%% file: paper3_v14_arxiv.tex
\documentclass[twocolumn]{aastex63}
\usepackage{graphicx}
\usepackage{amsmath}
\usepackage{enumitem}
\usepackage[normalem]{ulem}

\DeclareMathOperator*{\argmax}{arg\,max}

\shorttitle{Emission Signatures of SBHBs}
\shortauthors{Nguyen et al.}

\begin{document}
\def\redtext#1{#1}
\def\redtext#1{{\color{red}#1}}
\def\bluetext#1{#1}
\def\bluetext#1{{\color{blue}#1}}
\def\greentext#1{#1}
\def\greentext#1{{\color{green}#1}}
\title{Emission Signatures from Sub-parsec Binary Supermassive Black Holes III: Comparison of Models with Observations}


\correspondingauthor{Tamara Bogdanovi\'c}
\email{tamarab@gatech.edu, khainguyen@gatech.edu}

\author[0000-0003-3792-7494]{Khai Nguyen}
\affiliation{Center for Relativistic Astrophysics, School of Physics, Georgia Institute of Technology, Atlanta, GA 30332, USA}

\author[0000-0002-7835-7814]{Tamara Bogdanovi\'c}
\affiliation{Center for Relativistic Astrophysics, School of Physics, Georgia Institute of Technology, Atlanta, GA 30332, USA}

\author[0000-0001-8557-2822]{Jessie C. Runnoe}
\affiliation{Department of Astronomy, University of Michigan, 1085 S. University Avenue, Ann Arbor, MI 48109}
\affiliation{Department of Physics \& Astronomy, Vanderbilt University, 6301 Stevenson Center Ln, Nashville, TN 37235, USA}

\author[0000-0002-3719-940X]{Michael Eracleous}
\affiliation{Department of Astronomy \& Astrophysics and Institute for Gravitation and the Cosmos, Pennsylvania State University\\
525 Davey Lab, University Park, PA 16802}

\author[0000-0002-8187-1144]{Steinn Sigurdsson}
\affiliation{Department of Astronomy \& Astrophysics and Institute for Gravitation and the Cosmos, Pennsylvania State University\\
525 Davey Lab, University Park, PA 16802}

\author[0000-0001-9481-1805]{Todd Boroson}
\affiliation{Las Cumbres Observatory Global Telescope Network, Goleta, CA 93117}


\begin{abstract}
We present a method for comparing the H$\beta$ emission-line profiles of observed supermassive black hole (SBHB) candidates and models of sub-parsec SBHBs in circumbinary disks. Using the approach based on principal component analysis we infer the values of the binary parameters for the spectroscopic SBHB candidates and evaluate the parameter degeneracies, representative of the uncertainties intrinsic to such measurements. We find that as a population, the SBHB candidates favor the average value of the semimajor axis corresponding to $\log(a/M) \approx 4.20\pm 0.42$ and comparable mass ratios, $q>0.5$. If the SBHB candidates considered are true binaries, this result would suggest that there is a physical process that allows initially unequal mass systems to evolve toward comparable mass ratios (e.g., accretion that occurs preferentially onto the smaller of the black holes) or point to some, yet unspecified, selection bias.
Our method also indicates that the SBHB candidates equally favor configurations in which the mini-disks are coplanar or misaligned with the binary orbital plane. If confirmed for true SBHBs, this finding would indicate the presence of a physical mechanism that maintains misalignment of the mini-disks down to sub-parsec binary separations (e.g., precession driven by gravitational torques). The probability distributions of the SBHB parameters inferred for the observed SBHB candidates and our control group of AGNs are statistically indistinguishable, implying that this method can in principle be used to interpret the observed emission-line profiles once a sample of confirmed SBHBs is available but cannot be used as a conclusive test of binarity.

\end{abstract}

\keywords{galaxies: active --- galaxies: nuclei ---  methods: analytical --- quasars: emission lines}

\section{Introduction}\label{sec:intro}

Over the past decade spectroscopic searches have identified about a hundred supermassive black hole binary (SBHB) candidates at sub-parsec orbital separations. These searches rely on detection of the Doppler shift in the emission-line spectrum of an active galactic nucleus (AGN) that arises as a consequence of SBHB orbital motion, under assumption that at least one of its constituent SBHs can shine as an AGN. In this context, the Doppler-shifted broad emission lines (BELs) have been interpreted as originating in gas that is gravitationally bound to the individual supermassive black holes \citep[SBHs; e.g.][]{gaskell83, bogdanovic09, shen10}. The main complication of this approach however, is that the Doppler shift signature is not unique to SBHBs and can be mimicked by AGNs powered by single SBHs \citep[e.g.,][]{popovic12}. For example, \citet{barth15} find that in AGNs with unshifted line profiles, which are not known to be hosts to SBHBs, the centroid of the lines can fluctuate by $\sim 200-300\, {\rm km\,s^{-1}}$ on timescales of the order of the light crossing time of the broad-line region (BLR), just as a result of the reverberation of light. 

If any of the SBHB candidates targeted by the spectroscopic searches are true binaries, they are expected to have orbital periods $\sim {\rm few}\times10^{1-2}$ years \citep{pflueger18}. This indicates that sustained, multi-year follow-up observations, carried out by multiple groups, may soon be able to definitively identify the signatures of orbital motion in some candidates \citep{bon12, bon16, eracleous12, decarli13, ju13, liu13, shen13, runnoe15, runnoe17, li16, wang17,guo19}. 

Using more than one observational technique provides opportunities for SBHB detection as well as additional means to test their nature. For example, some SBHBs that are targeted by spectroscopic surveys may in principle also be detected with direct imaging of double nuclei using very long baseline interferometry \citep{rodriguez06, bansal17, dorazio18}, by detection of quasi-periodicity in their light curves \citep{valtonen08,graham15, charisi16, liu16, zheng16}, or by detection of signatures of a circumbinary disk cavity in AGN spectral energy distributions \citep[e.g.,][]{gultekin12, roedig14b, foord17, guo20}. Assuming that this or a similar approach produces a set of confirmed SBHBs, it is of interest to determine what can be learned about this population of objects from their observational signatures. This question is in the focus of this and two earlier papers of the same series that investigate the spectroscopic signatures from sub-parsec binary SBHs. 
 
In \citet{nguyen16}, hereafter Paper~I, we introduced a semi-analytic model to calculate the BEL profiles emitted from circumbinary accretion flows around sub-parsec SBHBs\footnote{Throughout this paper we refer to them as {\it modeled} or {\it synthetic} profiles.}. We modeled SBHB accretion flows as a set of three accretion disks: two mini-disks that are gravitationally bound to the individual black holes and a circumbinary disk. We neglected contribution to the flux of the narrow streams of gas that flow from the inner edge of the circumbinary disk to the mini-disks. The line luminosity of the streams, produced by photoionization, depends on the solid angle they subtend to the ionizing source. Since this solid angle is relatively small, the contribution of the streams to the emission-line flux is expected to be small. 
Given a physically motivated parameter space occupied by the sub-parsec binaries, we calculated a synthetic database of nearly 15 million BEL profiles and explored the dependence of the profile shapes on characteristic properties of the SBHBs. We have found that the model profiles in the first generation database show distinct statistical properties as a function of the binary semimajor axis, mass ratio, eccentricity, and the degree of alignment of the three disks.  A central result of that paper is that the BELs can in principle be used to infer the distribution of these physical parameters. That finding provided initial indication that diagnostic power of BEL profiles merits further investigation.

In \cite{nguyen19}, hereafter Paper~II, we presented the improved, second generation model and profile database by including the effect of radiation-driven accretion disk wind on properties of the BEL profiles, and by increasing the number of synthetic profiles to about 42.5 million. Under the influence of an accretion disk wind the emission-line profiles appear narrower, more symmetric, and predominantly single-peaked. The properties of such profiles are in better agreement with those of the observed sample of SBHB candidates and AGNs in general. Prior to implementing this effect, the database of modeled profiles presented in Paper~I contained more diverse profile morphologies (significant fraction of which had multiple peaks) and on average broader profiles than the observed SBHB candidates or a general population of AGNs.

 Analysis of the second generation database showed that correlations between the properties of the profiles and SBHB physical parameters identified in Paper~I are preserved, indicating that their diagnostic power is not diminished. That paper reports that the profile shapes are a more sensitive measure of the binary orbital separation and the degree of alignment of the black hole mini-disks and are less sensitive to the SBHB mass ratio and orbital eccentricity. By performing a preliminary comparison, based on profile distribution functions, we found that model profile shapes are more compatible with our observed sample of SBHB candidates than with the control sample of regular AGNs\footnote{The observed datasets are published in \citet{eracleous12} and \citet{runnoe15, runnoe17} and we refer to them as the E12 sample, search or dataset hereafter. See \S~\ref{sec:description} for more detail.}. Furthermore, that early comparison suggested that if the observed sample of SBHBs is made of genuine binaries, it must include compact systems with comparable masses and misaligned mini-disks. 

In this, third paper of the series we present a method for comparison of the model and observed optical BELs, based on principal component analysis, and use it to infer the properties of 88 SBHB candidates from the E12 spectroscopic search. The new aspect of this method is that in addition to the parameter estimates it also provides a quantitative measure of the parameter degeneracy, thus allowing us to establish uncertainties intrinsic to such measurements. This paper is organized as follows. We describe the method used to infer physical parameters for the observed SBHB candidates in \S\,\ref{sec:methods} and present the results for individual SBHB candidates and the entire sample in \S\,\ref{sec:results}. In \S\,\ref{sec:discussion} we discuss the implications of these results along with the limitations of our method and present our conclusions in \S\,\ref{sec:conclusions}.

\section{Methods}\label{sec:methods}

\subsection{Model for emission-line profiles from an SBHB in a circumbinary disk}

In Papers~I and II we presented a database of BEL profiles associated with gas accretion flows surrounding gravitationally bound SBHBs. We consider sub-parsec SBHBs with total mass $M=M_1+M_2$ and mass ratios $q = M_2/M_1$, where $M_1$ and $M_2$ are the mass of the primary and secondary black holes, respectively. We do not explicitly adopt a value for the SBHB mass, because the relevant properties and results of our calculation scale with this parameter (e.g., any length scales and the monochromatic emission-line flux). The results are nevertheless valid for a range of masses that correspond to black holes powering regular, nonbinary, AGNs (i.e., $\sim 10^6 - 10^9\,M_\odot$). The values of the key parameters of the SBHB in circumbinary disk model are listed in Table~\ref{table:parameters}. The full list of parameters of the model and their definitions can be found in Papers~I and II.

The binary orbits are characterized by a range of separations given by the orbital semimajor axis, $a$, expressed in units of $M\equiv GM/c^2 = 1.48\times10^{13}\,{\rm cm}\, (M/10^8\,M_\odot)$, where we use the binary mass as a measure of length in geometric units with $G=c=1$. For example, for a total SBHB mass of $M=10^8\,M_\odot$, the adopted range of semimajor axes shown in Table~\ref{table:parameters} corresponds to binary separations of $a \sim 0.025 - 5$\,pc. SBHBs are placed on either circular or eccentric orbits, described by the orbital eccentricity, $e$. The orientation of the observer relative to the SBHB orbit is given by the inclination angle, $i$, where $i = 0^\circ$ represents a clockwise binary seen face-on, and values $i> 90^\circ$ represent counterclockwise binaries (see Figure 17 of Paper~I for illustration of geometry).

\begin{deluxetable}{ll}
\tabletypesize{\normalsize}
\tablecolumns{4}
\tablewidth{0pt} 
\tablecaption{Key Parameters of the SBHB in Circumbinary Disk Model}\label{table:parameters}
\tablehead{Parameter & Value}
\startdata
$q$ & 1 , 9/11 , 2/3 , 3/7 , 1/3 , 1/10 \\
$a/M$ & $5\times 10^3$ , $10^4$ , $5 \times 10^4$, $10^5$, $10^6$ \\
$e$ & 0.0 , 0.5 \\
$i$ & $5^{\circ}$, $55^{\circ}$, $105^{\circ}$, $155^{\circ}$\\
$\theta_1$, $\theta_2$ & $0^{\circ}$, $30^{\circ}$, $60^{\circ}, 105^{\circ}$, $135^{\circ}$, $165^{\circ} $\\
$\tau_0$ &  0 ($10^{-4}$) , 0.1 , 1 , $10^2$ \\
\enddata
\tablecomments{$q$ -- SBHB mass ratio. $a$ -- Orbital semimajor axis. $e$ -- Orbital eccentricity. $i$ -- Inclination of the observer relative to the SBHB orbital angular momentum. $\theta_i$ -- Inclination of the primary and secondary mini-disk relative to the SBHB orbital angular momentum. $\tau_0$ -- Optical depth parameter.}
\label{table:parameters}
\end{deluxetable}

The accretion flow around the SBHB is described as a set of three circular, Keplerian accretion disks: two mini-disks that are gravitationally bound to their individual SBHs, and a circumbinary disk. The three disks are modeled as independent BLRs, where the size of the two mini-disks and the central opening in the circumbinary disk are constrained by the size of the binary orbit and are subject to tidal truncation by the SBHB, as described in Paper~I.  Therefore, the outer radii of the two mini-disks are always smaller than the binary separation, and the outer radius of the circumbinary disk is chosen to be $3 a$. For some model configurations this implies BLRs that are larger or smaller than $\sim 0.1-1$\,pc, the range usually found in luminous AGNs.

We do not eliminate any configurations based on the sizes of their BLRs, since these are not known for BLRs that may exist in binary SBHs. Rather, we compare the emission-line profiles of all model configurations to the observed ones, and reject those that produce profiles inconsistent with observations. In the context of this model, contributions to the flux from the three disks are summed into a resulting, composite emission-line profile. As described in Paper~II, because of their proximity to the two AGNs, the contribution to the profile flux is dominated by the mini-disks and the contribution from the circumbinary disk is negligibly small in majority of configurations.

We assume that the circumbinary disk is coplanar with the binary orbital plane and relax any assumptions about the orientation of the mini-disks.  The orientation of the mini-disks is described by the angles $\theta_1$ and $\theta_2$ and is measured relative to the angular momentum vector of the binary. For example, when $\theta_1 = \theta_2 = 0^\circ$, both mini-disks are coplanar with the SBHB orbit, and when $\theta_i > 90^\circ$, the gas in the mini-disks exhibits retrograde motion relative to the circumbinary disk. This setup provides a variety of configurations in which the three disks are illuminated by the two AGNs at different incidence angles. 

As noted in Papers~I and II, this characteristic illumination pattern results in emission-line properties distinct from those in regular AGN, where a single source of illumination is located in the center of the BLR. The disk misalignment can also lead to the ``shielding" of one AGN by the mini-disk associated with the companion SBH, as seen from the perspective of a distant observer. We account for this effect when such configurations arise by allowing the blocked AGN to illuminate its own mini-disk and the circumbinary disk but not the mini-disk of the other SBH. However, we do not take into account the eclipse of one disk by another, which can arise in misaligned configurations.

In Paper~II, we build on this model and present an improved database of emission-line profiles by taking into account the effect of the radiation-driven accretion disk wind occurring in the accretion flow surrounding the SBHB. Similarly to regular AGNs, we assume that the origin of the line-driven wind is the inner accretion disk of each SBH ($r \sim10^{14}$\,cm for $\sim10^8\,M_\odot$ SBHs), where dense gas blocks the soft X-ray photons from the compact source of continuum radiation but transmits UV photons, which allows radiation pressure on resonance lines to accelerate the outflow to $\sim0.1$c \citep{mc95}. The wind extends to larger radii in each disk, where it affects the structure and kinematics of the gas in the BLR (see Figure~1 in Paper~II for illustration of geometry of the BLR affected by the accretion disk wind).

We explore this phenomenon in the context of the low-ionization H$\beta$ lines emerging from the BLRs surrounding SBHBs, but the same calculation is in principle applicable to other emission lines. We assume the H$\beta$ emission region to be a very thin layer on the surface of the outer accretion disk, which in AGNs extends from $\sim10^{15}$ to $\sim10^{18}$\,cm in the radial direction. Specifically, the emission region resides at the interface between the disk and the wind where the gas is starting to accelerate. As a result, there is a significant velocity gradient in this layer but the radial and vertical velocity of the gas is negligible. Before escaping to infinity, some H$\beta$ emission-line photons are absorbed by a low-density accretion disk wind. The wind has a finite optical depth in the optical Balmer emission lines, thus modifying the intensity and shape of the emitted profiles \citep{cm96,flohic12}.  

The probability that the H$\beta$ line photons escape the wind can be estimated as a function of the local 
optical depth to line emission, calculated along the observer's line of sight. In Paper~II, the effect of line optical depth at a given radius in the disk is quantified by a normalization constant, $\tau_0$, a parameter which encapsulates the properties of the disk (its density, opacity and turbulent velocity) at the inner edge of the BLR. Note that $\tau_0$ does not express the true optical depth; in fact, the true optical depth is a function of azimuth and radius and can vary by several orders of magnitude at different positions in the emission layer. Throughout the manuscript we refer to $\tau_0$ as the optical depth parameter. In this work, the emission-line profiles are calculated for a range of optical depth parameters, $\tau_0 = 0$ to $10^2$, as shown in Table~\ref{table:parameters}. Because the profiles calculated with $\tau_0 = 0$ and $10^{-4}$ are very similar, we use them interchangeably. We have also verified that profile shapes remain unchanged for $\tau_0 > 100$, and we do not explore the values of optical depth parameter beyond this threshold. We refer the reader to section~2 of Paper~II for more detailed description of the remaining disk wind parameters.

\subsection{Description of the database of modeled and observed emission-line profiles}\label{sec:description}

Using the model outlined in the previous section we have calculated a database containing about 42.5 million profiles, associated with different SBHB configurations, as well as with different orientations of the observer relative to the binary. Of these, 18 million profiles are calculated for binaries on circular and 24.5 million on eccentric orbits. Similarly, 15 million profiles are calculated with $\tau_0 = 0$ and the rest have non-zero values of disk wind optical depth parameter. The profiles are divided approximately proportionally across the remaining key parameters of the model, shown in Table~\ref{table:parameters}. The entire synthetic database and associated open source scripts can be accessed at \url{https://github.com/bbhpsu/synthetic_spectra}. 

Based on the analysis of the modeled profile database presented in Paper~II, we have found that radiative transfer in the disk wind affects the overall shape of emission-line profiles by making them narrower on average and more symmetric in SBHB systems characterized by low $q$ and $i$. The shapes of modeled emission-line profiles are a sensitive function of the binary orbital separation and the degree of alignment in the triple-disk system but tend to be less sensitive (or more degenerate with respect) to the SBHB mass ratio and orbital inclination relative to the observer. Because there is a large degree of overlap between the models of SBHBs on circular and eccentric orbits, we do not expect that the profile shapes alone can be employed as a useful diagnostic of eccentricity. These earlier findings guide our expectations, in terms of the diagnostic power of the emission-line profiles and degeneracy of the SBHB parameters, as we set out to quantify them in this work.

We compare the database of synthetic profiles to the emission lines observed and published as a part of the E12 search for sub-parsec SBHBs. The E12 campaign searched for $z < 0.7$ Sloan Digital Sky Survey AGNs \citep[DR7;][]{schneider10}, with broad H$\beta$ lines offset from the rest frame of the host galaxy by $\gtrsim 1000\,{\rm km\,s^{-1}}$. Based on this criterion, E12 selected 88 quasars for observational follow-up from an initial group of about 15,900 objects. The radial velocity shifts of the H$\beta$ lines in these objects are so large ($\sim1000 - 6000\,{\rm km\,s^{-1}}$) that they cannot be explained by fluctuations due to the reverberation of light seen in AGNs with unshifted profiles, as these are smaller by an order of magnitude \citep{barth15, guo19}. This makes them a compelling candidate sample on which to test predictions for physical phenomena other than reverberation, including SBH binaries\footnote{As noted in Section~4.3 of E12, a fraction of objects have profiles that are ``boxy" or flat-topped, although still shifted, reminiscent of disk-like line profiles \citep[$\sim 25$\%;][]{eh94, eh03, strateva03}. A smaller fraction of the profiles have weak shoulders opposite the offset peak, reminiscent of some of the variable disk-like profiles \citep[$\sim 7$\%;][]{gezari07}. These are some of the reasons why we regard the quasars in this sample as candidates. A more detailed discussion of caveats is given in Section~6.3 of E12.}. This sample has a median redshift of 0.32, with a full range of $0.077 < z < 0.713$. The monochromatic luminosity distribution spans the range $43.27 < \log{[\lambda\,L_\lambda (5100\,\AA)/{\rm erg\,s^{-1}}]} < 44.68$, with a median value of 43.75 and a standard deviation of 0.31 \citep[see][for detailed description of the SBHB candidate sample]{runnoe15}.

The follow-up observations of the sample of SBHB candidates span a temporal baseline from a few weeks to 12\,yr in the observer's frame. Their goal is to measure the epoch-to-epoch variation in the velocity offset of the H$\beta$ profiles and to test the binary hypothesis by ruling out any sources whose radial velocity curve is not consistent with SBHB orbital motion for reasonable physical properties. After multiple epochs of follow-up, reliable measurements of radial velocity curves were obtained for 29/88 candidates (some of which show no significant velocity variations) and reported by the E12 campaign. 

At the present time, this approach has highlighted several promising cases for further follow-up but has not yet led to ruling out the SBHB hypothesis for any candidates. The epoch-to-epoch profile variability of the SBHB candidates on months-to-years timescales is of the order $\sim10 - 300\,{\rm km\,s^{-1}}$, and could in principle be a consequence of the reverberation of light, superposed with a physical phenomenon that caused the large absolute offsets of their profiles. We use a data set of broad optical emission lines (drawn from the E12 data set), which at the time of this analysis included 330 multi-epoch spectra of 88 SBHB candidates and 527 spectra for a control sample of 212 matching regular AGNs with similar redshifts and luminosities. 

The redshifts of the comparison sample span the range $0.08 < z < 0.68$, with a median value of 0.38, very similar to the binary candidates. The monochromatic luminosity distribution spans the range $43.44 < \log{[\lambda\,L_\lambda (5100\,\AA)/{\rm erg\,s^{-1}}]} < 45.17$, with a median value of 43.88 and a standard deviation of 0.33 \citep[see][for detailed description of the control sample]{runnoe15}. It is worth keeping in mind that while the 212 control sample AGNs were not targeted as binary hosts, the presence of SBHBs among them cannot be ruled out. We thus use them for comparison under the hypothesis that they are {\it plausible} nonbinary AGNs. The dataset containing the observed spectra of SBHB candidates and a control group of AGN is available at \url{https://github.com/bbhpsu/spectra}.

In order to isolate the broad H$\beta$ line and quasar continuum, \citet{runnoe15} performed a spectral decomposition that deblends the quasar continuum, optical Fe~$\textsc{ii}$, H$\beta$, and [O~$\textsc{iii}$] emission components (see \S~2.3 of their paper for detailed description of this procedure). In the final step, they decomposed the H$\beta$ profiles using Gaussian components: two to characterize the narrow H$\beta$ and two for the broad H$\beta$ line. Occasionally, five Gaussians were used in decomposition of very complex profiles (two for the narrow and three for the broad line). The analysis described in this paper uses the parametric reconstruction of the broad component of the H$\beta$ line obtained from this procedure\footnote{Throughout this work we refer to this dataset as {\it observed profiles} or {\it observed sample}, for simplicity.}. This allows us to make a direct comparison with the synthetic broad H$\beta$ profiles produced by our model, which by design do not include other emission components or noise.

\subsection{Comparison of the modeled and observed samples using principal component analysis}\label{sec:comparisson}

\begin{figure*}[]
\centering
\includegraphics[width=0.85\textwidth, clip=true]{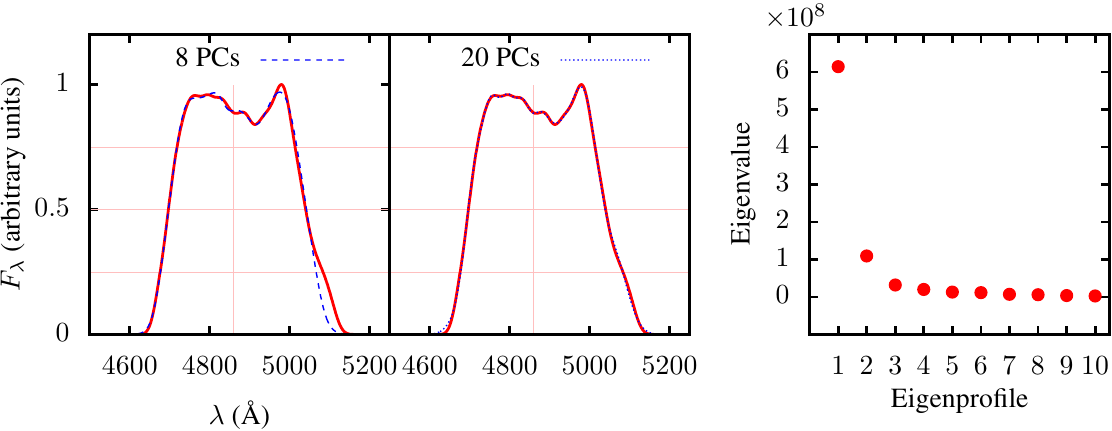}
\caption{Illustration of the reconstruction of a profile from the modeled database (marked by the red solid line) using principal component analysis. The left panel shows reconstruction of the profile using the first 8 and the first 20 principal components (blue dashed and dotted lines, respectively). The right panel shows the variance of the synthetic database along each principal axis, given by the eigenvalues, as a function of the order of the eigenprofile. Even though the first 8 principal components represent more than 98 percent of the total variance of the modeled database, a reliable reconstruction of more complex profiles, like the one shown, requires about 20 principal components. The vertical red line marks the rest wavelength of the H$\beta$ line}.
\label{fig:reconstruction}
\end{figure*}

The analysis carried out in Papers~I and II has unambiguously showed that SBHB properties are imprinted in the population of the modeled BEL profiles, albeit with some degeneracy. It has also provided a statistical statement about the collective properties of the observed SBHB candidate sample but did not provide the means to determine the parameters of individual binary candidates. We have therefore developed a method, based on principal component analysis (PCA), which allows us to infer the properties of individual SBHB candidates, as well as to quantify the uncertainties associated with those determinations. 

PCA allows decomposition of a dataset into a number of linearly independent principal components, or eigenvectors. In the case of a sample that consists of emission-line profiles, the eigenvectors are represented by eigenprofiles. The leading (``zeroth") component of the decomposition is the average spectrum and higher-order components (first, second, etc.) represent progressively less likely modifications of the average spectrum needed to reproduce a given profile. This technique is optimal for analysis of large and complex data sets, which cannot be inspected visually. For example, \citet{bl09} extended the application of this technique to spectra of 9,800 SDSS quasars with the goal of identifying outliers among them. The same method has been used to flag the SBHB candidates for observational followup from the spectra of $\sim15,900$ SDSS quasars in the E12 sample.  

 In this work, we use PCA to decompose the sample of synthetic and observed emission-line profiles using the same basis of eigenprofiles. The basis is derived from the the synthetic spectra and is then used to represent the observed spectra. This approach allows us to perform a comparison of the two samples in order to reveal the portion of the parameter space favored by the SBHB candidates. The eigenvalue that corresponds to each eigenprofile is a measure of its relative importance in accounting for the variance within the sample. In this scheme, the highest weight goes to the defining features, present in the majority of the profile sample and lower weight goes to the features in which noise or a unique profile characteristic dominates. The procedure used to compute eigenprofiles is described in Appendix~\ref{sec:Calc}.

Figure~\ref{fig:reconstruction}, which shows PCA reconstruction of one of the most complex BEL profiles selected from our database. Most of BELs in our database have simpler profiles and can be successfully reconstructed with only 8 eigenprofiles. This is a reflection of the fact that the eigenvalues of the first 8 eigenprofiles represent more than 98 percent of the total variance of the modeled database (the first 8 eigenprofiles are illustrated in Figure~\ref{fig:eigenspectra} in Appendix~\ref{sec:Calc}). In the case of the profile shown in Figure~\ref{fig:reconstruction}, the first eight eigenprofiles do not fully describe the profile features, and a faithful reconstruction of this profile (and other profiles of similar complexity) requires at least about 20 eigenprofiles. Consequently, we choose the first 20 eigenprofiles to ensure that we can accurately describe all synthetic profiles in our database. The same set of eigenprofiles used for description of the synthetic database is then used to describe the profiles in the observed SBHB candidate sample and the control sample of AGNs.

\begin{figure*}[t]
\centering
\includegraphics[width=0.9\textwidth, clip=true]{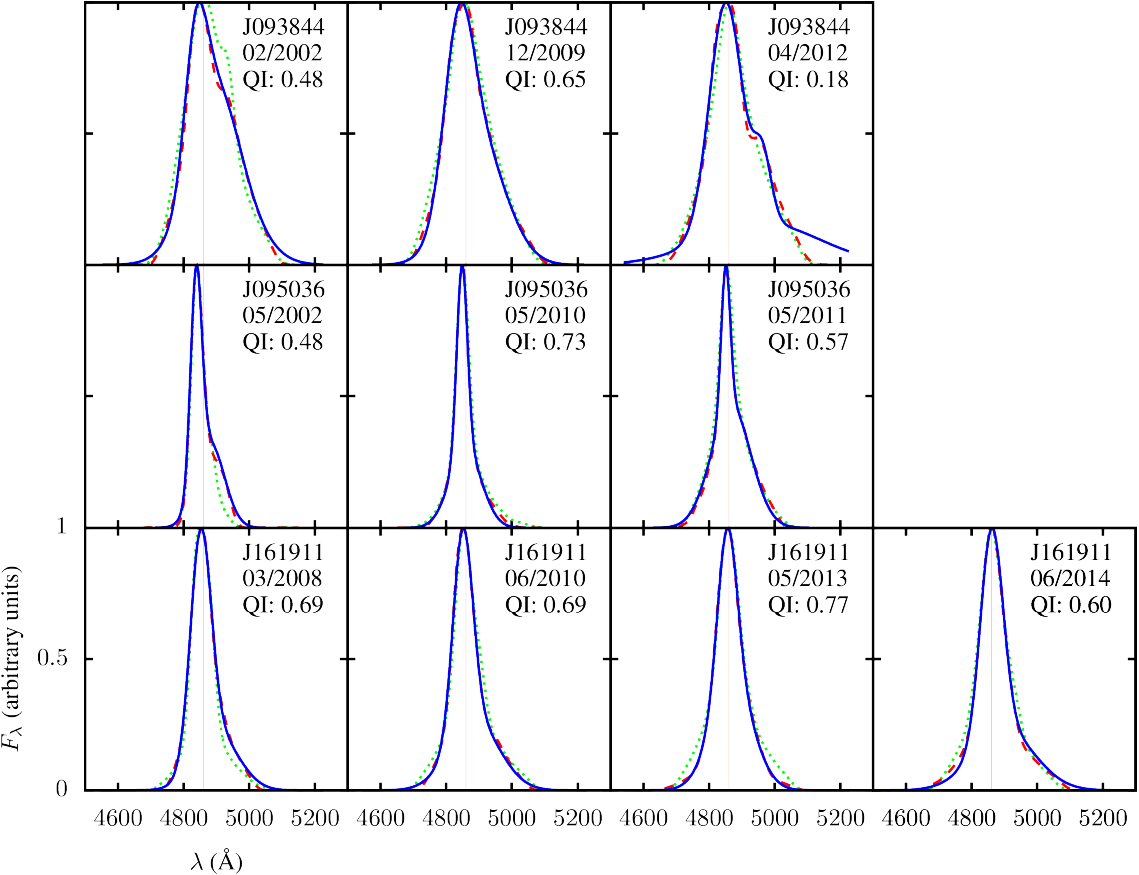}
\caption{H$\beta$ emission-line profiles from multi-epoch observations of three SBHB candidates: SDSS J093844 (top row), J095036 (middle), and J161911 (bottom). For each epoch, we show the observed profile (blue solid line), the synthetic profile with the smallest (red dashed) and largest distance (green dotted) from the observed profile, contained in the nearest neighbor set. The vertical red line marks the rest wavelength of the H$\beta$ line.}
\label{fig:cutoff}
\end{figure*}
%
\begin{figure*}[]
\centering
\includegraphics[width=0.9\textwidth, clip=true]{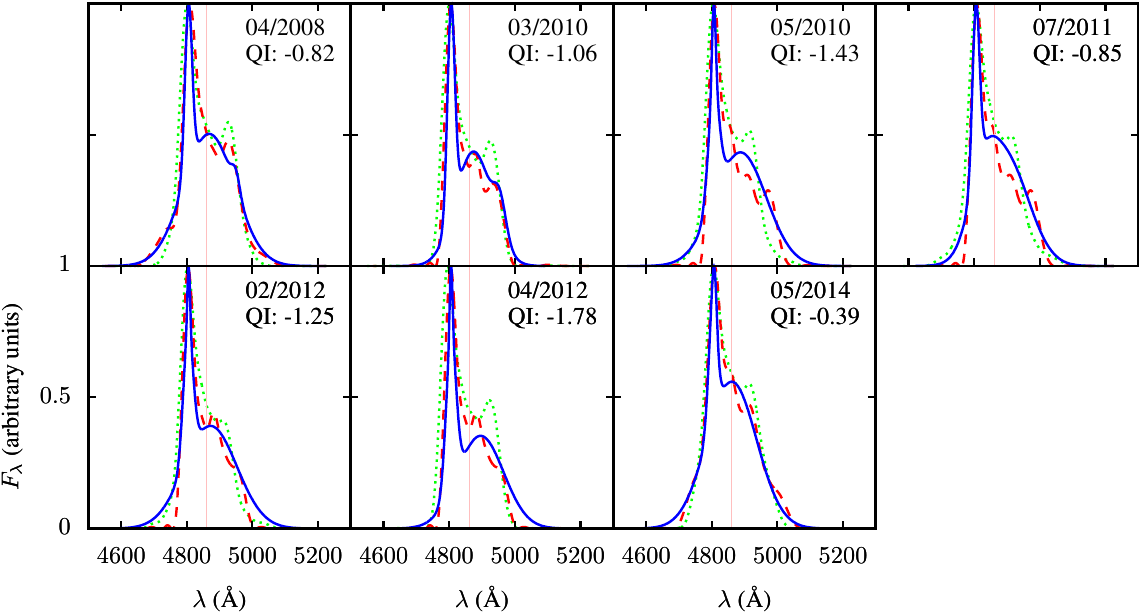}
\caption{H$\beta$ emission-line profiles from multi-epoch observations of the SBHB candidate J153636, indicated by the time stamp. The observed candidate has the lowest average QI among all objects from the E12 sample of SBHB candidates, indicating that there is no close match for its profiles in the synthetic database. For each epoch, we show the observed profile (blue solid line), the synthetic profile with the smallest (red dashed) and largest distance (green dotted) from the observed profile, contained in the nearest neighbor set. The vertical red line marks the rest wavelength of the H$\beta$ line.}
\label{fig:worstQI}
\end{figure*}

Having defined a common set of basis vectors for both datasets, we use it to compare the modeled and observed profiles by calculating their Euclidean distance in the space defined by the 20 eigenprofiles. We define a distance from a given observed profile ($\mathbf{F}^o$) to an arbitrary profile in the synthetic database ($\mathbf{F}^s$) as
\begin{equation}
d(\mathbf{F}^s,\mathbf{F}^o) = \left[ \sum_{i=1}^{20} (T^s_i-T^o_i)^2  \right]^{1/2}
\label{eq:distance}
\end{equation}      
Here, $\mathbf{F}^o$ and $\mathbf{F}^s$ are vectors of size $\left[1 \times \mathrm{M}\right]$ and $\mathrm{M}=600$ is the number of equal frequency bins used to describe each profile. $\mathbf{T}^s$ and $\mathbf{T}^o$ contain the weights assigned to the constituent eigenprofiles, used to reconstruct the synthetic and observed profile, respectively (see Appendix~\ref{sec:Calc}). We rank all synthetic profiles in terms of their distance from the observed profile to obtain the synthetic database sorted by distance, $\tilde{\mathbf{F}}$, such that, $d(\tilde{\mathbf{F}}^s,\mathbf{F}^o)\leq d(\tilde{\mathbf{F}}^{s+1},\mathbf{F}^o)$.

From the ranked database we select a subset of $\mathcal{N}$ profiles that are the nearest neighbors to the observed profile, whose number is determined as the larger of the number of profiles within some cutoff distance $d_{\textrm{c}}$ and 6500.
\begin{equation}
\mathcal{N}(\mathbf{F}^o)=\max\left[k:d(\tilde{\mathbf{F}}^k,\mathbf{F}^o)<d_{\textrm{c}}(\mathbf{F}^o), \,6500\right]
\label{eq:Ksize}
\end{equation}
The cutoff distance $d_{\textrm{c}}$ is chosen to be within 10 percent of the ``modulus" of the observed profile, 
\begin{equation}
d_{\textrm{c}}(\mathbf{F}^o) = 0.1 \left[ \sum_{m=1}^\mathrm{M} \left(F^o_m\right)^2 \right]^{1/2} \,.
\label{eq_dc}
\end{equation}
Here, $F^o_m$ represents the monochromatic profile flux at a given wavelength. The cutoff distance therefore sets the margin for variation in the profile shape when searching the synthetic database for nearest neighbors. The value of 10 percent is chosen arbitrarily, so to enclose only a portion of the synthetic database, while still ensuring a statistically significant number of neighbor profiles for majority of the observed profiles. The minimum number of neighbor profiles used in our analysis is $k=6500$. This value corresponds approximately to the square root of the total number of the synthetic profiles in our database. This is a common choice in algorithms used to sort the data, since it allows the number of steps, of the order $\mathcal{O}(k^2)$, to scale linearly with the size of the database.
 
Therefore, each observed profile has a well-defined set of nearest neighbors in the synthetic database and a cutoff distance calculated using equation~\ref{eq_dc}.  Figure~\ref{fig:cutoff} shows a visual comparison between the multi-epoch profiles for three observed SBHB candidates (SDSS~J093844, J095036, and J161911) and the profiles with the smallest (most similar) and largest distance (least similar) belonging to their corresponding nearest neighbor sets. In a majority of cases, the nearest and furthest neighbor defined in this way are similar in shape to the observed profiles. The exception are the cases in which the synthetic database does not contain a profile similar enough to the observed profile, as shown for the spectrum of J093844 observed in April 2012.

In order to measure the quality of the achieved match, we define the quality index (QI) as a function of the distance between the observed profile and its closest neighbor in the synthetic database
\begin{equation}
\textrm{QI} = \frac{d_c-d_{\rm min}}{d_c} \,\,.
\end{equation}  
By definition, $0\leq {\rm QI} \leq 1$ for observed profiles whose closest neighbor can be identified within its corresponding cutoff distance. In the cases when the closest neighbor cannot be found within the cutoff distance, the algorithm by default selects the closest 6500 synthetic profiles as its nearest neighbors. Such scenarios result in ${\rm QI} <0$, indicating a lower quality match, since the distance to the nearest synthetic profile $d_{\rm min} > d_{\textrm{c}}$.  We illustrate this case in Figure~\ref{fig:worstQI}, which shows a sequence of seven observed BEL profiles for the candidate J153636. QI for this object remains negative for every epoch of observation. Furthermore, its average quality index, ${\rm QI} = -1.08$ (calculated as a simple average for all epochs of observation), is the lowest in the entire database, indicating that interpretation of this SBHB candidate is less reliable, simply because there is no close match for it in the synthetic database. The fraction of SBHB candidates with a negative average QI makes up about 18\% of the E12 sample and we list their values in Appendix~\ref{sec:ParamTable}.

 \subsection{Calculation of probability distributions for inferred SBHB parameters}\label{sec:distributions}

The procedure described in the previous section allows us to determine a set of synthetic profiles, which are the closest neighbors to each observed profile. Since every synthetic profile corresponds to a unique set of SBHB parameters, we use the set of nearest neighbor profiles to map each observed profile into a preferred portion of the SBHB parameter space. In this approach, the average value of the SBHB parameters associated with a group of the nearest neighbor profiles represents a binary configuration favored by the observed profile, and the variance in the value of each SBHB parameter provides a measure of its degeneracy.

The SBHB parameters favored by the observed profile are calculated as a weighted average of the values associated with its nearest neighbor profiles, in such a way that the synthetic profiles closer to the observed profile contribute more to the average. The weight for each nearest neighbor profile is defined as an exponential function of its distance 
\begin{equation}
w(\mathbf{F}^s,\mathbf{F}^o)=\begin{cases} 
	e^{-5\,d/d_{\textrm{c}}} \,, &  \{a, q, e, i \} \,\, \textrm{repeat}\\
	0 \,, &  \{a, q, e, i \} \,\, \textrm{do not repeat}
\end{cases}
\label{eq:weight}
\end{equation}
The kernel in equation~\ref{eq:weight} puts most of the weight on profiles within $d_{\textrm{c}}$ and a negligible weight on profiles at larger distances. We confirmed that the exact form of the weight function has a weak impact on the results of our analysis, as long as it decreases rapidly with distance.

The first condition in equation~\ref{eq:weight} requires that four physical parameters of the SBHB model, namely the $a$, $q$, $e$, and $i$, are repeated in every epoch of observation of the same binary candidate. This requirement is based on the expectation that the binary separation, mass ratio, eccentricity, as well as the orientation of its orbital plane are unlikely to change significantly from one epoch of observation to another, which for the E12 monitoring campaign corresponds to $\lesssim 12$ years. Other parameters of the model, such as the optical depth parameter of the disk wind, orientation of the mini-disks, etc., are not subject to this constraint. Note that the mini-disks are not necessarily expected to precess on such short timescales either but we chose not to constrain their orientation. Even with 42.5 million profiles in the database, the sampling of the parameter space is not sufficiently dense to allow explicit constraints on many parameters of the model. As we show in \S~\ref{sec:results}, this still allows us to calculate probability distributions for the mini-disk orientations, albeit with a greater degree of degeneracy.

For example, the H$\beta$ emission-line profiles associated with the SBHB candidate J095036 have been observed in three different epochs (see Figure~\ref{fig:cutoff}). The three observed profiles have three corresponding sets of synthetic profiles, each containing $\mathcal{N}_1$, $\mathcal{N}_2$, and $\mathcal{N}_3$ nearest neighbor profiles, as determined by equation \ref{eq:Ksize}. If a synthetic profile from the second set has a combination of parameters $\{a,q,e,i\}$, that is repeated in some of the profiles contained in the set one and three, these profiles are assigned a non-zero weight, according to equation~\ref{eq:weight}. On the other hand, the synthetic profiles whose combination of parameters is not represented in all three nearest neighbor sets simultaneously are assigned zero weight. Therefore, a requirement that multi-epoch observations should map into the same portion of the $\{a,q,e,i\}$ parameter space allows us to further constrain the SBHB parameters and to reduce their degree of degeneracy.
 
Once the nearest neighbors of an observed profile and their weights are determined, the probability distribution for a given SBHB parameter can be calculated as
\begin{equation}
\textrm{Pr}(x=x') = \frac{\sum_{s=1}^{\mathcal{N}}w(\mathbf{F}^s,\mathbf{F}^o):x(\mathbf{F}^s)=x'}{\sum_{s=1}^{\mathcal{N}}w(\mathbf{F}^s,\mathbf{F}^o)} \,,
\label{eq:pdf}
\end{equation}
where $x$ represents an SBHB parameter of interest and $\textrm{Pr}$ is a discrete probability density function (PDF) when $x=x'$. This PDF is a multivariate function and $x$ represents a vector of all physical parameters of the model. For example, fixing the value of one of the parameters, $\textrm{Pr}(a=5000M)$ is equal to the sum of the weights for all nearest neighbors of an observed profile, such that their combination of parameters includes $a=5000M$, and is normalized by the sum of the weights. This procedure is repeated for all values of $a$ and the corresponding $\textrm{Pr}(a)$ calculated for every epoch of observation in which an SBHB candidate is observed. The resulting, multi-epoch PDF is calculated as a simple average of PDFs from all epochs of observation. Finally, equipped with a PDF for every SBHB parameter, we can calculate the mean value of each parameter and its standard deviation, for every binary candidate.

In addition to the mean and standard deviation, we also calculate the ``entropy" and use it as a common statistical measure of degeneracy of each inferred SBHB parameter
\begin{equation}
\mathcal{S}_x= -\sum_{j=1}^{\mathrm{N}_{\rm ch}} \frac{ \textrm{Pr}(x=x_j) \log \left(\textrm{Pr}(x=x_j) \right)  } {\log\left(\mathrm{N}_{\rm ch}\right)} \,\,.
\label{eq:entropy}
\end{equation}
Here, $\mathrm{N}_\textrm{ch}$ denotes the number of parameter choices as shown in Table~\ref{table:parameters}, for instance $\mathrm{N}_\textrm{ch}(a)=5$. According to equation~\ref{eq:entropy}, a well-defined PDF with no degeneracy corresponds $\mathcal{S}=0$, and a maximally degenerate, uniform PDF corresponds to $\mathcal{S}=1$. The range of values for entropy defined in this way allows us to compare the degeneracy of different SBHB parameters on the same scale. 


\section{Results}\label{sec:results}


\subsection{Analysis of individual SBHB candidates}\label{sec:individual}

\begin{figure*}[]
\centering
\includegraphics[width=1.0\textwidth, clip=true]{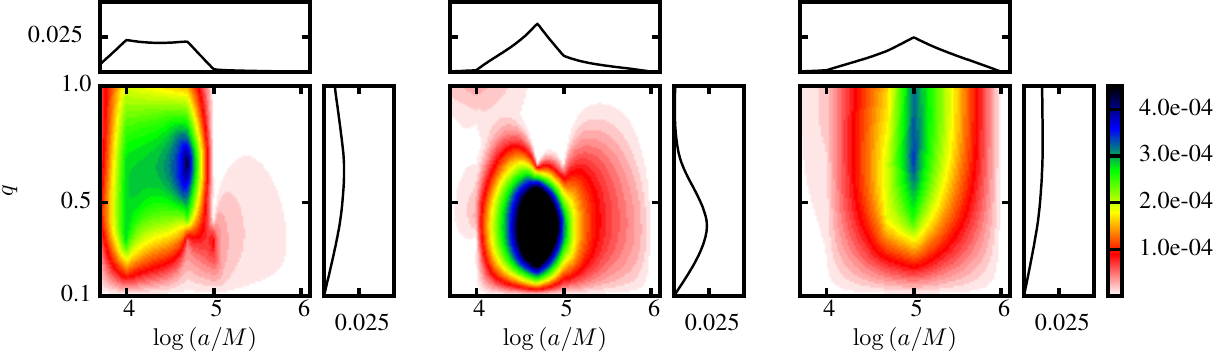}
\caption{2D probability density distribution in terms of $\log(a/M)$ and $q$ for SDSS~J093844 (left), J095036 (middle), and J161911 (right panel). The rectangular insets show the 1D projections.}
\label{fig:individualaq}
\end{figure*}
%
\begin{figure*}[]
\centering
\includegraphics[width=1.0\textwidth, clip=true]{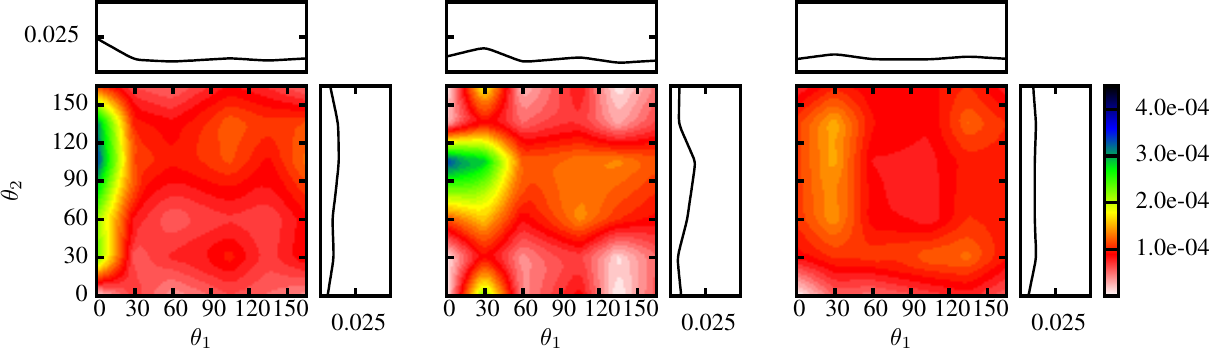}
\caption{2D probability density distribution in terms of $\theta_1$ and $\theta_2$ for SDSS~J093844 (left), J095036 (middle), and J161911 (right panel). The rectangular insets show the 1D projections.}
\label{fig:individualt1t2}
\end{figure*}
%
\begin{figure*}[]
\centering
\includegraphics[width=1.0\textwidth, clip=true]{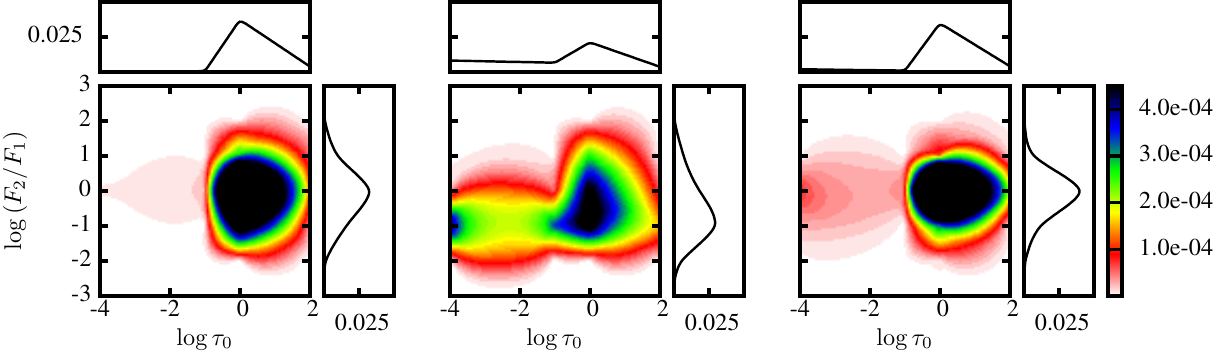}
\caption{2D probability density distribution in terms of $\log \tau_0$ and $\log{(F_2/F_1)}$ for SDSS~J093844 (left), J095036 (middle), and J161911 (right panel). The rectangular insets show the 1D projections.}
\label{fig:individualtauF2F1}
\end{figure*}

In this section, we present the analysis of three SBHB candidates from the E12 sample and note that the same analysis has been carried out on the remaining group of the SBHB candidates and control AGNs. The objects SDSS~J093844, J095036, and J161911 are of interest since they have been highlighted by \citet{runnoe17} as the most promising SBHB candidates in the sample, based on the properties of their radial velocity curves. 
More specifically, the radial velocity curves of these candidates show a statistically significant monotonic change in radial velocity over the duration of observations. As shown in Figure~\ref{fig:cutoff}, the nearest neighbor profiles in the synthetic database provide a good description of the observed profiles for J093844, J095036, and J161911, as reflected by their average quality indices, $\rm QI = 0.44$, 0.59 and 0.69, respectively. 

Figure~\ref{fig:individualaq} shows the PDFs for the three SBHB candidates in terms of $a$ and $q$. The 2D distributions are made by dividing the parameter space into $100\times 100$ equal bins and by interpolating the probability in each bin from the discrete PDF obtained via equation~\ref{eq:pdf}. The interpolated PDFs are then renormalized to one, resulting in values of $\sim10^{-4}$. Visual inspection shows that the favored values of semimajor axes for all three SBHB candidates fall in the range of $\sim 10^4-10^5\,M$. Appendix~\ref{sec:ParamTable} lists their mean values for $\log(a/M)$ and the associated standard deviations as $(4.19 \pm 0.34)$, $(4.64 \pm 0.28)$, and $(4.81 \pm 0.26)$,  for J093844, J095036, and J161911, respectively. Similarly, the values of $q$ for the same three candidates are $(0.65 \pm 0.22)$, $(0.43 \pm 0.17)$, $(0.72 \pm 0.22)$. 

Among the three candidates, J095036 (middle) is characterized by the smallest degree of degeneracy in the inferred values of $\log(a/M)$ and $q$, as witnessed by their values of entropy listed in Appendix~\ref{sec:ParamTable} ($\mathcal{S}_a \approx \mathcal{S}_q = 0.44$). This is because the properties of the profiles of J095036 (such as the location of the peak and profile asymmetry), show significant change from one observation to another. This epoch-to-epoch variability provides an effective way to reduce the SBHB parameter degeneracies, as it helps to eliminate parameters which values are not repeated in all epochs of observations (see equation~\ref{eq:weight}). 

Along similar lines, in the entire sample of the SBHB candidates, J131945 (listed as number 56 in Appendix~\ref{sec:ParamTable}) has the best constrained value of the semimajor axis. This candidate is a good example of potential gains provided by the continued spectroscopic monitoring of SBHB candidates: it has been observed 9 times between April 2002 and April 2013, whereas most of the other objects in the sample have 3 to 4 observations and a similar baseline. This, combined with the fact that the broad base of its H$\beta$ profiles shows significant variability from one observation to another, guarantees that very few SBHB configurations can produce all of the observed profiles. It is important to note however that the observed profile variability may be driven by some process in a single (or indeed isolated) BLR, not captured in our SBHB model. This adds ambiguity to the interpretation of observed profiles, as our model does not distinguish variability due to such processes from those related to the binary phenomenon.

Figure~\ref{fig:individualt1t2} shows the PDFs for the SBHB candidates, J093844, J095036, and J161911, in terms of the angles $\theta_{1}$ and $\theta_{2}$ (the distributions have been calculated in the same way as those in Figure~\ref{fig:individualaq}). The two angles are of interest because they describe the orientations of the two mini-disks relative to the binary orbital plane. As noted earlier, the emission-line profiles presented in this work have been calculated assuming that both accreting SBHs can shine as AGNs and illuminate their own mini-disk, as well as the two other disks in the system. The effect of illumination of one mini-disk by a companion AGN is however most pronounced in binaries when their mini-disks are misaligned with the SBHB orbital plane. This nonaxisymmetric illumination pattern by the two AGNs can give rise to very asymmetric profiles whose shapes can vary on timescales shorter than the SBHB orbital period. Conversely, such profiles can in principle be a sensitive probe of their alignment.

Figure~\ref{fig:individualt1t2} however indicates that the three SBHB candidates under consideration have no strongly preferred values for $\theta_{1}$ and $\theta_{2}$. More specifically, while there is a weak preference for $\theta_1 \sim 0^\circ$ for candidate J093844 and $\theta_2 \sim 100^\circ$ for J095036, their 1D distributions show a significant degree of degeneracy. This large degree of degeneracy can in part be explained by the fact that we do not impose the requirement that the values of $\theta_{1}$ and $\theta_{2}$ must be repeated in every epoch of observation, as is done for the other parameters in equation~\ref{eq:weight}. 
 Better constraints on the mini-disk orientations can in principle be obtained, by more sophisticated modeling of $\theta_{1}$ and $\theta_{2}$ as a function of the binary orbital phase.

Figure~\ref{fig:individualtauF2F1} shows the PDFs for the same three SBHB candidates in terms of the optical depth parameter, $\tau_0$, and the ratio of the H$\beta$ emission-line flux contributed by the secondary and primary mini-disk, $F_2/F_1$. The value of the optical depth parameter is relatively well constrained for J093844 (left panel) and J161911 (right) and it peaks at  $\tau_0\approx 1$. The same property is more degenerate for J095036 (middle), which, as noted earlier, is the candidate with the narrowest emission-line profiles among the three. This can be understood as the disk wind has less effect on the shapes of narrower profiles, and hence,  accretion disk wind with a range of optical depths can produce profile shapes similar to the relatively narrow profiles of J095036. The degree of degeneracy in the inferred value of $\tau_0$ can also be inferred from the value of the entropy calculated for this parameter, which amounts to $\mathcal{S}_{\tau_0} = 0.27$, $0.85$ and $0.42$ for J093844, J095036, and J161911, respectively.

The property $\log(F_2/F_1)$ shown in Figure~\ref{fig:individualtauF2F1} is calculated from our model (i.e., it is not an input parameter) and is of interest because it indicates which mini-disk dominates the H$\beta$ emission. In our model, the flux ratio is determined by three effects: $(i)$ the accretion rates onto the primary and secondary SBH, which are assumed to power the AGN emission in the UV and X-ray band, $(ii)$ the surface area of the mini-disks that are emitting the BEL profiles and $(iii)$ by the cross-illumination of the mini-disks by the companion AGN.

\begin{figure*}[]
\centering
\includegraphics[width=0.85\textwidth, clip=true]{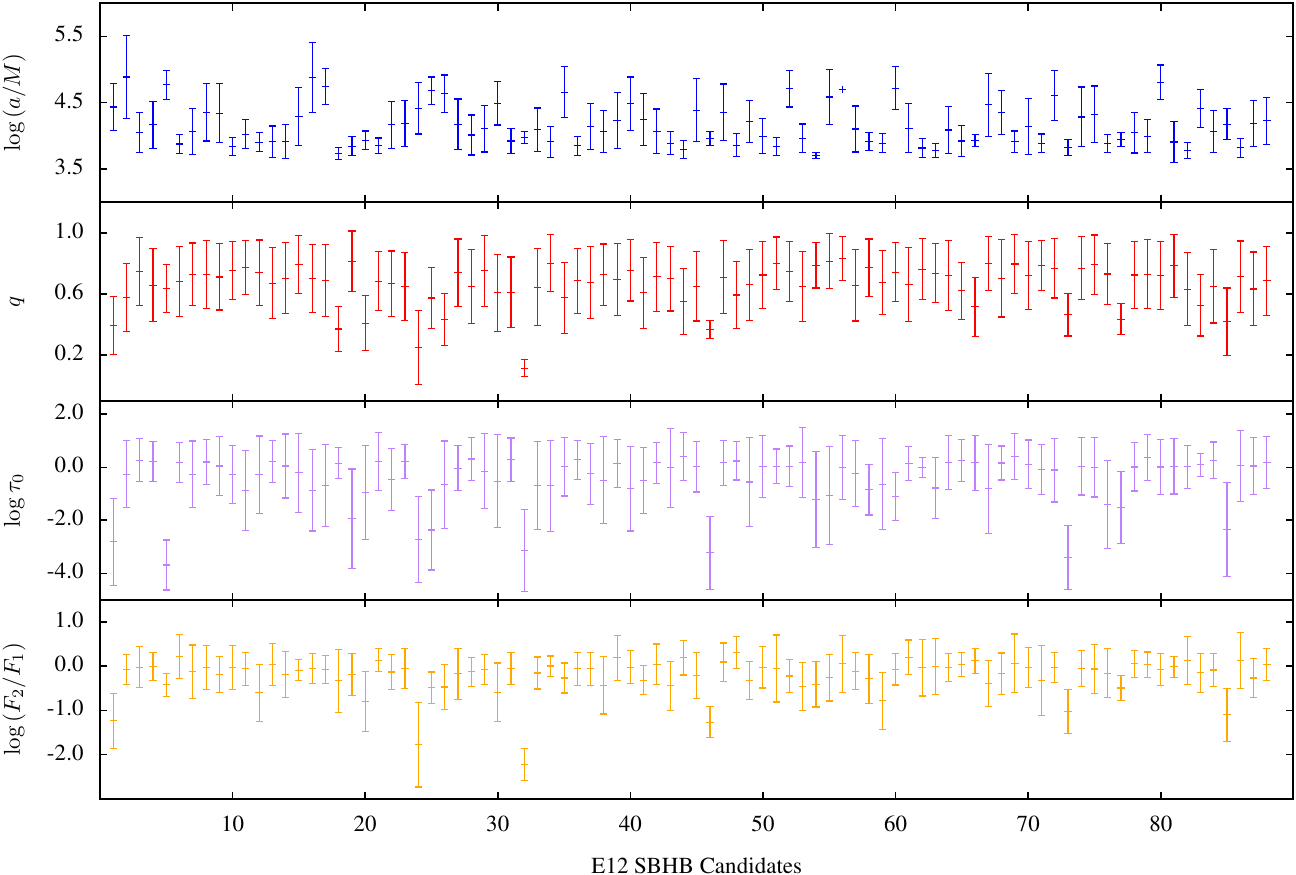}
\caption{Inferred average values of binary parameters and their standard deviations for 88 SBHB candidates from the E12 survey. From top to bottom: $a$, $q$, $\log \tau_0$ and $\log(F_2/F_1)$. Data corresponding to each panel can be found in Appendix~\ref{sec:ParamTable}.}
\label{fig:meanstd}
\end{figure*}

The effects $(i)$ and $(ii)$ are competing because the ratio of mass accretion rates onto the two SBHs, $\dot{M}_2/\dot{M}_1$, decreases with increasing $q$ (see equation~3 of Paper~I), whereas the ratio of the surface areas of their mini-disks increases approximately as $\sim q^2$. Since the latter effect dominates, we expect the two mini-disks to make comparable contributions to the emission-line flux when $q\sim 1$ and to have $\log(F_2/F_1) < 0$ when $q < 1$. Note that the latter expectation is different from the behavior of isolated BLRs, where higher luminosity isolated AGNs usually have more luminous broad emission lines. Unlike isolated BLRs, the mini-disks in our model are truncated by gravitational torques. Therefore, even though the AGN associated with a lower mass SBH is more luminous by assumption, its truncated BLR makes a smaller contribution to the flux of the composite H$\beta$ line relative to the primary BLR. The effect $(iii)$ requires certain degree of geometric misalignment of the mini-disks. Consequently, the cross-illumination of the mini-disks by the companion AGN affects the shapes of a smaller fraction of profiles in our database (see \S~4.2 in Paper~II for more discussion). When present, however, this effect can result in $\log(F_2/F_1)$ that is either positive or negative, depending on the configuration.

From the three binary candidates considered in this section, J093844 (left) and J161911 (right) have relatively high inferred mass ratios, $q\approx 0.7$, and consequently, the fluxes emitted by the two mini-disks are comparable. In contrast, J095036 (middle) has $q\approx 0.4$ and a correspondingly lower peak value of $\log(F_2/F_1)\approx -0.5$. 

We also consider the ability of our method to infer the orbital eccentricity of an SBHB from its BEL profiles. With just two choices for this parameter in our synthetic database ($e=0$ and $e=0.5$), the SBHB orbital eccentricity of the observed candidates cannot be sufficiently constrained by our method. While this choice was made in order to produce a synthetic database of manageable size, it is actually not clear that expanding the database further, by adding configurations with other eccentricities, would lead to improved ability to predict eccentricity. This is because the analysis carried out in Papers~I and II already shows that there is a large degree of overlap in profile properties for $e=0$ and $e=0.5$ cases, indicating substantial degeneracy in this parameter. This can be seen in Figure~3 of Paper~II that shows the distribution of profile asymmetry and location of the peak for circular and eccentric SBHBs. The two distributions occupy a similar footprint and have similar features, making them difficult to distinguish. The same is true for other profile distributions examined in the first two papers of the series. We therefore conclude that distinguishing among different cases of orbital eccentricity would be difficult, even if our database contained many different values.

\begin{figure*}[]
\centering
\includegraphics[width=0.85\textwidth, clip=true]{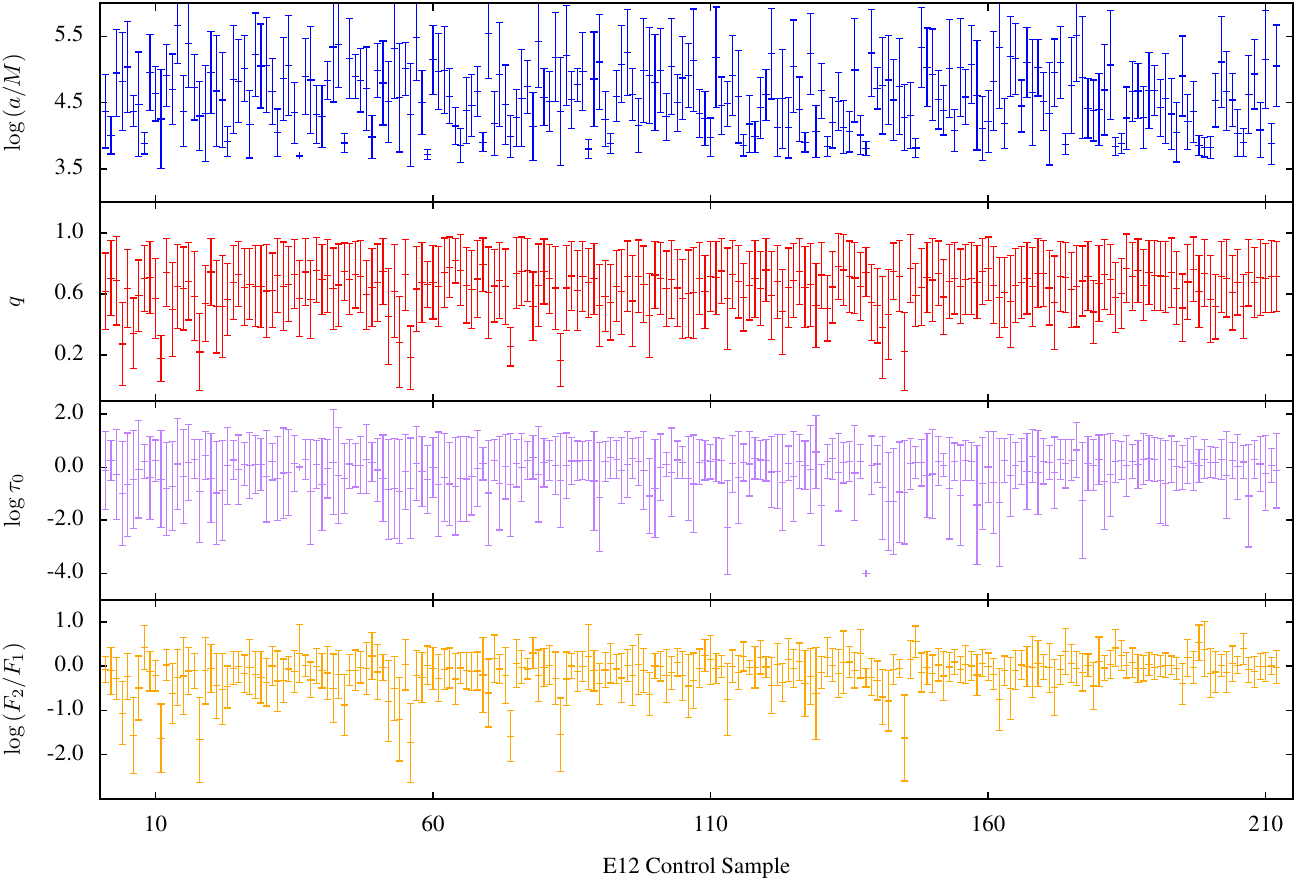}
\caption{Inferred average values of binary parameters and their standard deviations for  212 matching control sample AGNs from the E12 survey. From top to bottom: $a$, $q$, $\log \tau_0$ and $\log(F_2/F_1)$.}
\label{fig:AGNmeanstd}
\end{figure*}

Finally, we remind the reader that the values of the SBHB parameters inferred here are based on the analysis of the parametric representation of the broad component of the H$\beta$ line from \citet{runnoe15}, as described in \S~\ref{sec:description}. The result of this procedure are the broad H$\beta$ profiles that are smooth (i.e., without visible contribution from the spectral noise), as illustrated in Figures~\ref{fig:cutoff} and \ref{fig:worstQI}. Because modeling of the spectral noise is sidestepped in our analysis, the parameter values presented in this and the next section do not account for the impact of the spectral noise. We analyze this effect in Appendix~\ref{sec:NoisyProf}, by quantifying the uncertainty associated with the presence of the spectral noise in the data. This analysis indicates that the impact of the spectral noise in the E12 dataset on the values of the inferred SBHB parameters is smaller than the impact of the parameter degeneracy, represented by their standard deviation.

\subsection{Properties of the entire SBHB candidate sample}\label{sec:88candidates}

In this section we discuss the properties of the entire sample of 88 SBHB candidates from the E12 search. The top two panels of Figure~\ref{fig:meanstd} provide a visual summary of the mean values of $\log(a/M)$ and $q$ and their standard deviations for all 88 candidates. The preferred values of semimajor axes are similar to those of the three candidates discussed in the previous section. They range between $3.5 \lesssim \log(a/M) \lesssim 4.5$, or equivalently $0.015 \, \textrm{pc} \lesssim a \lesssim 0.15 \, \textrm{pc} $ for $M=10^8 M_{\odot}$ SBHB, and have standard deviations, $\sigma_a < 0.6$, where we adopt notation $\sigma_a = \sigma(\log(a/M))$ for brevity. In terms of the binary mass ratios, the values preferred by most candidates are in the range $0.2\lesssim q \lesssim 0.8$ and have standard deviations in the range $0.1 \lesssim \sigma_q \lesssim 0.3$. Furthermore, the corresponding values of entropy for most SBHB candidates tend to be higher for $q$ than for $a$, indicating a larger degree of degeneracy associated with the determination of the mass ratio.

The inferred values of the optical depth parameter cover a relatively wide range of values, $-3.4 \lesssim \log \tau_0 \lesssim 0.4$, seem to be moderately degenerate and characterized by the average value of entropy $\mathcal{S}_{\tau_0} \approx 0.5$. It is interesting to note that our database includes values of optical depth parameter as high as $\tau_0 = 100$ but no SBHB candidate favors the average value higher than a few. Visual inspection of Figure~\ref{fig:meanstd} shows that the systems with the lowest value of $\tau_0$ also tend to favor the lowest values of $q$.  Inspection of the emission-line profiles for these systems shows that they tend to be double-peaked and relatively smooth. Such profiles are well described by synthetic profiles in our database produced by systems in which the primary AGN is the dominant contributor to the emission-line flux and the optical depth of its disk wind is low, so that the double-peaked profile is preserved.

\begin{figure}[]
\centering
\includegraphics[width=0.45\textwidth, clip=true]{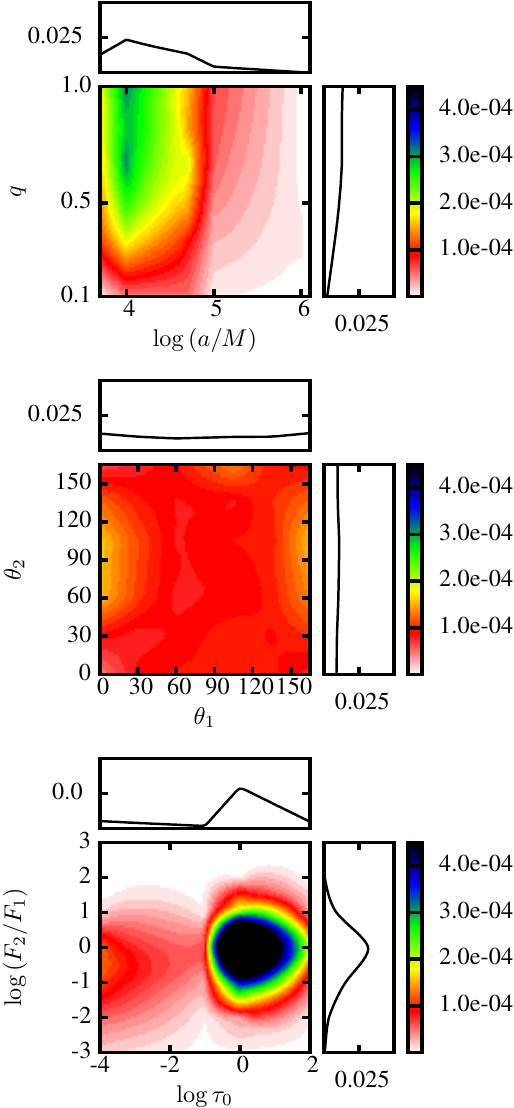}
\caption{2D probability density distributions in terms of $\log(a/M)$ and $q$ (top), $\theta_1$ and $\theta_2$ (middle), and $\log \tau_0$ and $\log{(F_2/F_1)}$ (bottom) for the 88 SBHB candidates from the E12 sample. The rectangular insets show the 1D projections.}
\label{fig:BBH}
\end{figure}

\begin{figure}[]
\centering
\includegraphics[width=0.45\textwidth, clip=true]{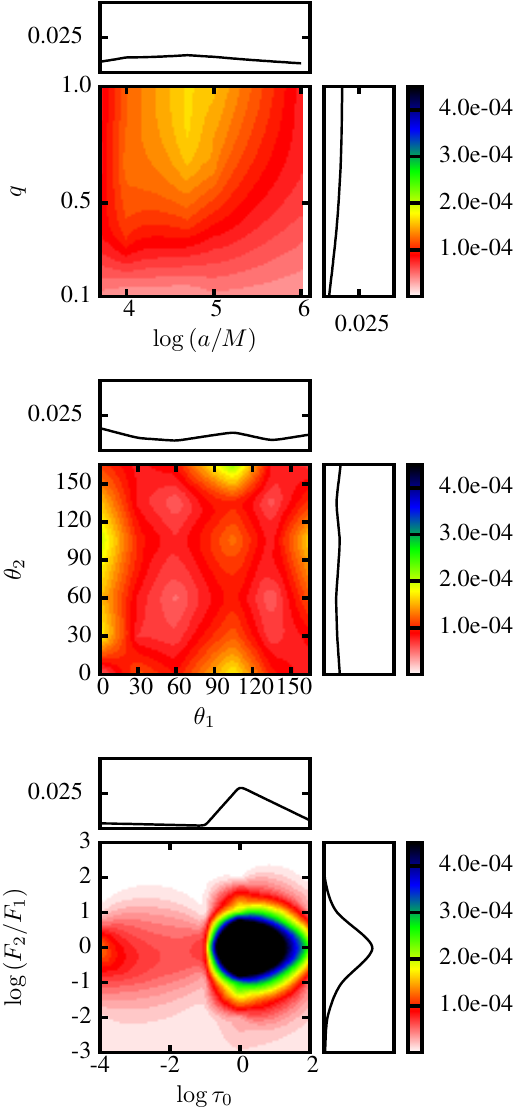}
\caption{Same as Figure~\ref{fig:BBH} for the control group of AGNs from the E12 sample.}
\label{fig:Control}
\end{figure}

Similarly, the inferred values of the flux ratio for most of the SBHB candidates correspond to comparable contributions to the line flux by the primary and secondary mini-disks. An exception to this are a few candidates with the flux ratio as low as $\log(F_2/F_1)\approx -2$, which also correspond to the systems with the lowest inferred values of mass ratio. This correlation is expected, based on the scaling of the mini-disk areas with the binary mass ratio, as explained in the previous section. The entropy for this parameter spans a wide range of values from one SBHB candidate to another, $0.06 \lesssim \mathcal{S}_{F_{2/1}} \lesssim 0.7$, indicating that the predictive power for the flux ratio varies a lot for different systems.

We perform the analysis of the spectra of the control group of 212 AGNs in the same way and show the resulting SBHB parameter values in Figure~\ref{fig:AGNmeanstd}. While there is no expectation that any (or nearly any) of the objects from this group host a binary, it is still interesting to consider whether they favor a different portion of the binary parameter space. The control group of AGNs favors a somewhat larger values of the semimajor axis, on average, and larger standard deviations, compared to the SBHB candidates. This can be understood because the control group of AGNs are on average characterized by profiles of smaller width (measured in terms of FWHM or FWQM; see Figure~9 in Paper~II and discussion therein). Unlike wider profiles, which tend to map into the SBHB configurations with larger orbital velocities and smaller separations, narrower profiles can also map into configurations with wider separations. Larger standard deviation in $\log(a/M)$ of the control group of AGNs is associated with the same effect. The values for the rest of the parameters however appear very similar to those inferred for the observed SBHB candidates.

\begin{figure*}[t]
\centering
\includegraphics[width=0.85\textwidth, clip=true]{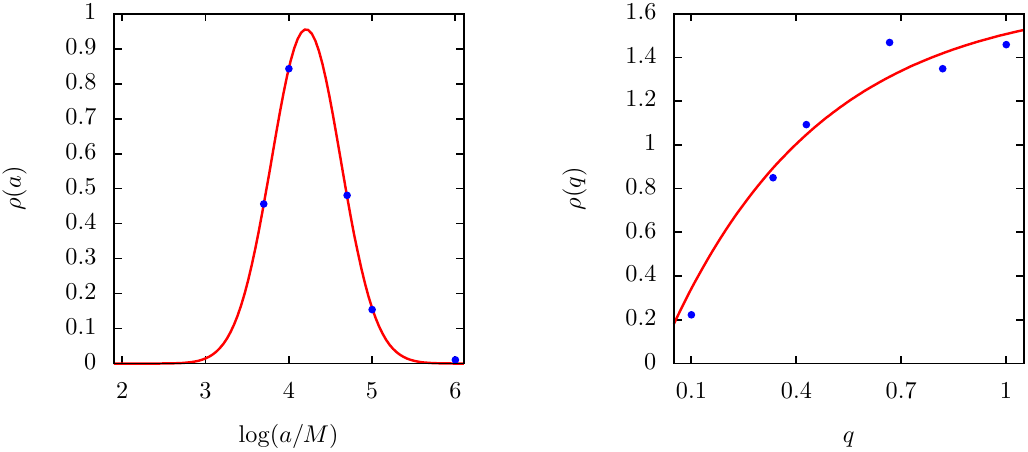}
\caption{Best fit for the continuous probability density function for all 88 SBHB candidates from the E12 sample in terms of the semimajor axis (left panel) and mass ratio (right). The blue dots mark the probability density determined by the comparison of the observed emission-line profiles from the SBHB candidates and synthetic profiles using our method.}
\label{fig:fitaq}
\end{figure*}

Figure~\ref{fig:BBH} shows the PDFs for the entire sample of 88 SBHB candidates from the E12 search. The distributions are calculated as the simple averages of distributions for individual candidates, equivalent to those shown in Figures~\ref{fig:individualaq} to \ref{fig:individualtauF2F1}. The top panel shows that as a population, the SBHB binary candidates favor the value of semimajor axis corresponding to $\log(a/M) \approx 4.20\pm 0.42$ and comparable mass ratios, $q > 0.5$. The width of this distribution is representative of those associated with individual SBHB candidates, even though they show a diversity of shapes on candidate-by-candidate basis, as shown in Figure~\ref{fig:individualaq}. 

SBHB candidates as a population show no strong preference for particular values of the angles $\theta_{1}$ and $\theta_{2}$. This is because individual SBHBs candidates have different combinations of $\theta_{1}$ and $\theta_{2}$ (see Figure~\ref{fig:individualt1t2}), which result in a relatively uniform average. The most interesting aspect of this statement is the implication that binary candidates do not seem to prefer the configuration in which the mini-disks are coplanar with the orbital plane more than any other configuration. The bottom panel of Figure~\ref{fig:BBH} captures the probability distributions very similar to those shown in the left and right panel of Figure~\ref{fig:individualtauF2F1}, with $\tau_0\approx 1$ and $F_2/F_1 \approx 1$.

Figure~\ref{fig:Control} shows the PDFs corresponding to the entire sample of 212 control AGN from the E12 search, as a comparison. As noted before, the control group of AGNs favors a somewhat larger average value of the semimajor axis and a larger standard deviation, $\log(a/M)\approx 4.60 \pm 0.72$, and the probability distributions for the remainder of the parameters are statistically indistinguishable. Specifically, neither control AGNs nor SBHB candidates show preference in terms of $\theta_1$ and $\theta_2$, and the favored values for $\tau_0$ and $F_2/F_1$ are the same for both groups.

This similarity indicates that the approach presented here can be used to infer the parameters once an SBHB candidate is confirmed as a real binary, but cannot be used as a test of binarity. It is worth noting that a preliminary comparison, reported in Paper~II, suggested that modeled profile shapes are more compatible with the observed sample of SBHB candidates than with the control sample of regular AGNs. That comparison was based on profile distribution functions (profile asymmetry, location of the peak, etc.), calculated for the synthetic and the two observed datasets. The comparison reported here is instead based on posterior distributions for SBHB parameters, which, as stated above, in many ways appear statistically indistinguishable. The two statements are not necessarily at odds -- they are just a different manifestation of the fact that mapping between the emission-line profiles and SBHB parameters in our model is characterized by some amount of degeneracy. In other words, even if the shapes of the profiles for the SBHB candidates and control AGNs appear statistically distinct, the distributions of the SBHB parameters inferred from them can overlap.

For practical purposes we also provide the analytic fits to the 1D distribution functions for the semi-major axis and the mass ratio for all 88 SBHB candidates. The continuous PDF for $\log(a/M)$ can be described by a normal distribution shown in the left panel of Figure~\ref{fig:fitaq} and hence, $\rho(a)$ can be expressed as
\begin{equation}
\rho (a) \propto \exp{ \left[ - \frac{\left[ \log (a/M)-4.2 \right]^2}{2 \times 0.42^2} \right] }, \,\,
3.7 \leq \log (a/M) \leq 6 .
\label{eqn:proba}
\end{equation}

In the same spirit, the continuous PDF for $q$ can be described with an exponential distribution\footnote{Alternatively, $q$ can also be described by $\propto q^{1/2}$ distribution.} shown in the right panel of Figure~\ref{fig:fitaq} and expressed as
\begin{equation}
\rho (q) \propto 1-\exp\left[-\frac{q}{0.44}\right], \qquad 0.1\leq q \leq 1
\label{eqn:probq}
\end{equation}

\section{Discussion}\label{sec:discussion}

\subsection{Implications for theory and observations}\label{sec:implications}

If the E12 sample of SBHB candidates were true binaries, their inferred PDFs would be a combination of the intrinsic properties of the binary (orbital separation, mass ratio, etc.) as well as the selection effects inherent to the spectroscopic search. The E12 and other spectroscopic searches are in principle sensitive to SBHBs with orbital separations in the range  $\sim 10^3 - 10^4\,M$ \citep[][P18 hereafter]{pflueger18}. The low and high end cutoffs for this range are set by two effects: (a) the binaries at smaller separations tend to evolve at a higher rate, making their detection less probable and (b) those at larger separations have radial velocity variations (determined from their BELs) that are too small to be detected by spectroscopic surveys on timescales of years. Taking these considerations into account we use the model developed by P18 to calculate the likelihood for detection of subparsec SBHBs given the parameters and selection effects of the E12 search. It is worth emphasizing that this likelihood and our interpretation of the SBHB candidates presented in Section~\ref{sec:results} are obtained independently, and therefore, their comparison provides a consistency check for the results obtained by the two methods. 

\begin{figure*}[t]
\centering
\includegraphics[width=0.9\textwidth, clip=true]{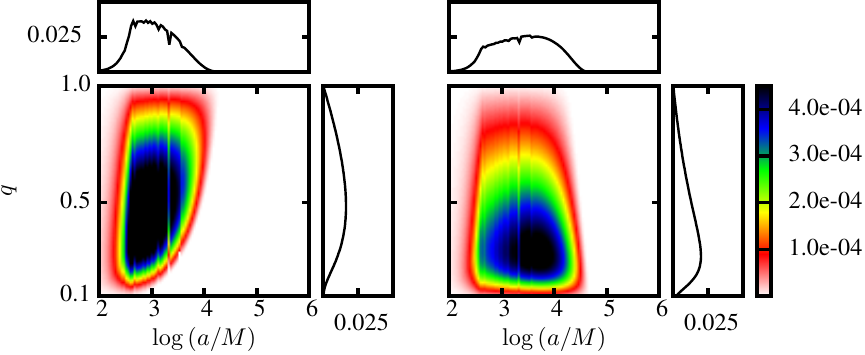}
\caption{Likelihood for detection of SBHBs given a yearly cadence of observations by the E12 spectroscopic search based on the P18 model of $10^7\,M_\odot$ binaries with an accretion rate through the circumbinary disk corresponding to $\dot{M} = 0.1 \dot{M}_E$. The left (right) panel illustrates the likelihood map for the SBHBs in which the primary (secondary) mini-disk make the dominant contribution to the flux of the H$\beta$ emission line. The color bar marks the normalization of the likelihood, which is arbitrary and chosen to match Figures~\ref{fig:BBH} and \ref{fig:Control}. An open source {\tt Python} script for calculation and plotting of the likelihood is available at \url{https://github.com/bbhpsu/Pflueger_etal18}.}
\label{fig:likelihood}
\end{figure*}

The panels of Figure~\ref{fig:likelihood} show the likelihood for detection of SBHBs given a yearly cadence of observations (comparable to the E12 spectroscopic search, on average), based on the P18 model of $10^7\,M_\odot$ binaries with the accretion rate through the circumbinary disk corresponding to $\dot{M} = 0.1 \dot{M}_E$. Here, $\dot{M}_E = L_E/\eta c^2$ is the Eddington accretion rate, $\eta$ is the radiative efficiency, $L_E = 4\pi GM m_pc/\sigma_T$ is the Eddington luminosity, $\sigma_T$ is the Thomson cross section, and other constants have their usual meaning. The left and right panel illustrate the likelihood map for the SBHBs in which the primary or the secondary mini-disk make the dominant contribution to the flux of the H$\beta$ emission line, respectively. 

In the case when the emission from the primary mini-disk dominates, there is a positive correlation between the mass ratio $q$ and the maximum semimajor axis that a detected SBHB can have. This is because as $q$ increases, the radial velocity due to the reflex motion of the primary SBH also increases. It follows that the binaries with larger mass ratios are favored in this case because they can be detected at larger orbital separations. Conversely, in the case when the emission from the secondary mini-disk dominates, the reflex motion of the secondary SBH is maximized for the smallest values of $q$. As a consequence, this scenario favors lower mass ratio binaries. Another difference worth pointing out is that the scenario when the primary mini-disk dominates places a stronger constraint on the binary semimajor axis, since in this case $a < 10^4\,M$, whereas $a < {\rm few}\times 10^4\,M$ when the secondary dominates, for parameters used in calculation of Figure~\ref{fig:likelihood}.

The analysis presented in Section~\ref{sec:results} indicates that most of the SBHB candidates in the E12 sample favor values of the mass ratio $q > 0.5$ and flux ratios $F_2/F_1 \lesssim 1$. A handful of remaining cases, with smaller values of $q$ typically have $F_2/F_1 \ll 1$. These values are consistent with the scenario in which the primary mini-disk dominates or makes contribution to the flux of the H$\beta$ emission-line comparable to the secondary mini-disk. This is of interest for two reasons. Firstly, it is contrary to the assumption commonly made by the spectroscopic surveys in the interpretation of their results. This assumption is directly motivated by a number of theoretical studies of SBHBs in circumbinary disks which show that in unequal-mass binaries accretion occurs preferentially onto the smaller of the two SBHs, which orbits closer to the inner edge of the circumbinary disk \citep{al96, gunther02,hayasaki07,roedig11,farris14}. 

Taken at face value, this suggests that the AGN associated with the secondary SBH may be more luminous than the primary. However, as noted earlier, the flux of the BELs is not merely determined by the bolometric luminosity of the AGN but also by the size of its BLR, which is in our model given by the surface areas of the two truncated mini-disks (the flux contribution by the circumbinary disk is small and can be neglected in all physically motivated configurations investigated by our model). Therefore, an important implication of our results for observational searches is that they should consider the case in which the measured radial velocity curves are associated with the primary SBH, and which would consequently point to more compact systems of SBHBs.

\begin{figure*}[t]
\centering
\includegraphics[width=0.9\textwidth, clip=true]{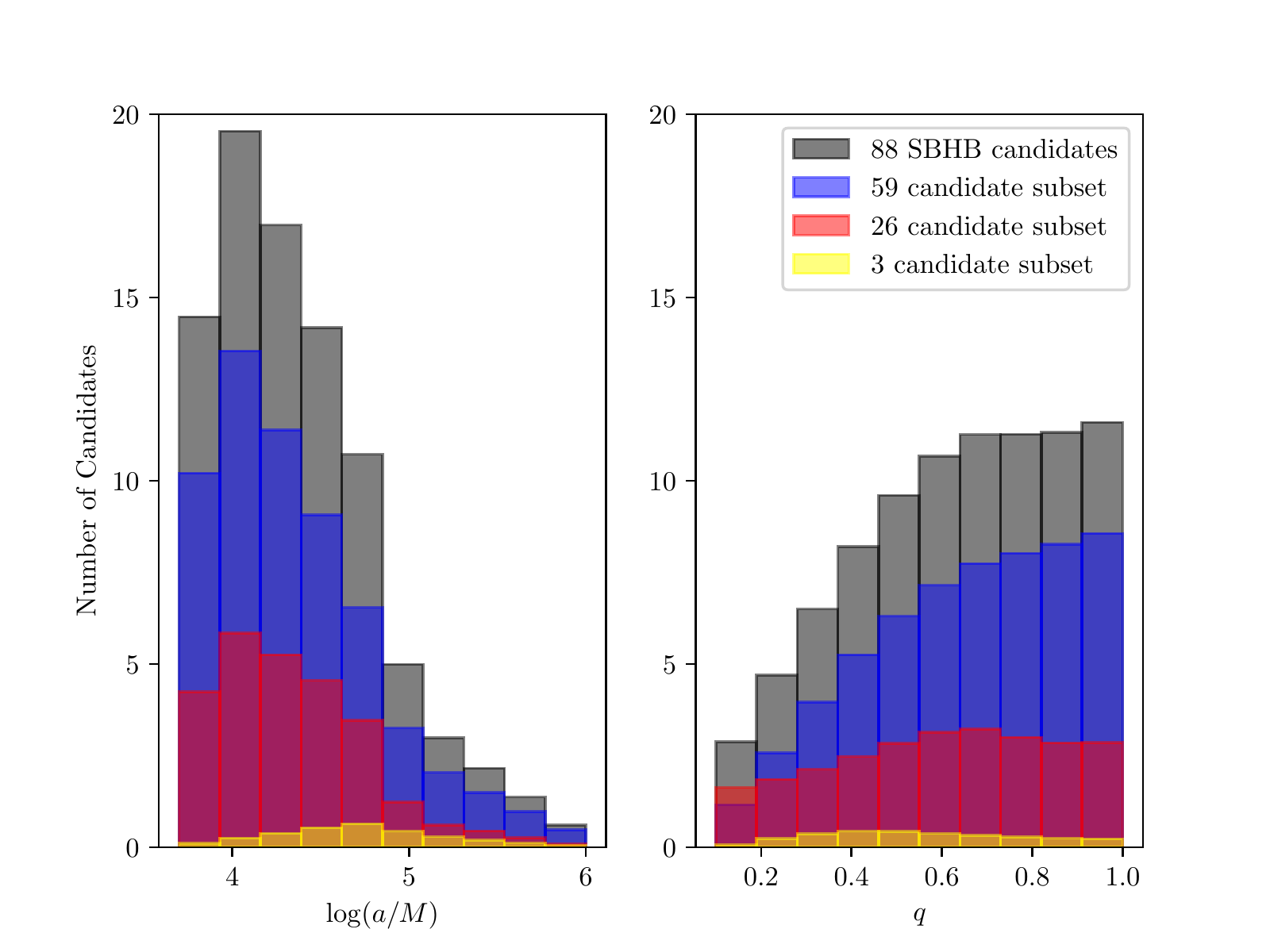}
\caption{Histograms showing the posterior distributions for $a$ (left) and $q$ (right) of 88 SBHB candidates, equivalent to those shown in Figure~\ref{fig:fitaq}. The candidates are divided into 3 subsets: 59 candidates for which radial velocity curve could not be measured, 26 candidates with radial velocity curve measurements, and 3 promising candidates discussed in \S~\ref{sec:individual} and in \citet{runnoe17}.}
\label{fig:histograms}
\end{figure*}

Secondly, the preference for the higher values of $q$ among the observed SBHB candidates also suggests that, if these are real binaries, there is a physical process that allows initially unequal mass systems to evolve toward comparable mass ratios. That scenario already seems to be borne out by the local simulations of SBHBs in circumnuclear and circumbinary disks, which show that accretion occurs preferentially onto the smaller of the SBHs. This presents an interesting challenge for cosmological models of binary evolution, which predict that sub-parsec SBHBs with lower mass ratios should be more abundant than those with comparable mass ratios \citep[e.g.,][see also P18 and references therein]{kelley17}. If true SBHBs indeed favor comparable mass ratios, this would suggest that the accretion rate inversion, reported by the local simulations of SBHBs in circumbinary disks, is an important ingredient that must be included in the cosmological models. Alternatively, it may point to some, yet unspecified, selection bias that favors detection of higher $q$ binaries.

Another result worth considering is that the observed SBHB candidates in our sample seem to equally favor configurations in which the mini-disks are aligned or misaligned (or warped) relative to the binary orbital plane. This is of interest because gravitational torques between the SBHs and the triple disk system can cause precession of the mini-disks, while diffusive processes can align the SBH spins and the mini-disk axes with the orbital axis \citep{miller13,hawley18}. If so, the alignment of the SBHB-spin-disk system is expected to evolve with binary separation, and the orientation of the mini-disks inferred from observations may be an important indicator of whether the mechanism leading to coplanar alignment is efficient.

It is worth noting that the E12 spectroscopic campaign has measured reliable radial velocity curves for 29/88 (some of which show no significant velocity variations) -- these are marked with ``1" in Appendix~\ref{sec:ParamTable}. The histograms with the posterior distributions for $a$ and $q$ for this subset of objects are shown in Figure~\ref{fig:histograms}, along with the rest of the candidate sample, for comparison. Of these, 3/29 candidates discussed in \S~\ref{sec:individual} were highlighted as the most promising SBHB candidates in the sample, based on the properties of their radial velocity curves. This is because their radial velocity curves show a statistically significant monotonic change in radial velocity over the duration of observations, as is expected from a binary that traced out some portion of its orbit around the center of mass.

E12 have not obtained measurements of the velocity modulation for 59/88  candidates whose profiles change in shape significantly from one epoch of observation to another (also shown in Figure~\ref{fig:histograms}). This is because significant changes in a BEL profile shape can either mimic or hide the change caused by the radial velocity modulation due to binary orbital motion, thus precluding a reliable measurement of the profile offset along the wavelength axis \citep[see Appendix~A in][for analysis of this effect]{runnoe17}. The method presented here is however particularly effective for SBHBs whose profile shapes change in time, because in these systems we obtain stronger constraints on the binary parameters, as discussed in Section~\ref{sec:results}. The two analyses therefore provide independent constraints complimentary to one another, because they infer SBHB properties from two different aspects of BEL profiles: their offset and their shape.

Figure~\ref{fig:histograms} shows that the three different subsets of SBHB candidates map into similar posterior distributions in $a$ and $q$, so at least according to these measures they seem to be objects of the same type. It is, of course entirely possible that none of the SBHB candidates in the E12 sample are actual binaries. If this can be shown, it would at the minimum indicate that SBHBs at sub-parsec separations produce no unique optical BEL signatures relative to the other known AGNs. This would raise a question whether the SBHB in the circumbinary disk model is the appropriate description for this class of binaries as, guided by theoretical models and simulations, one expects significant differences in the kinematics, geometry and photoionization properties of their BLRs, compared to the AGNs powered by single accreting SBHs. Even if they are not binaries, the 88 SBHB candidates represent unusual AGN specimens, whose long term monitoring may be an important step toward understanding the properties of the BLRs in general.


\subsection{Simplifications and limitations of the method}\label{sec:limitations}

Because the method presented here is built upon the first- and second-generation model and databases presented in Papers~I and II, the assumptions used there are also shared with this work. We direct the reader to Papers~I and II for detailed discussion of the implications of these simplifying assumptions and only address the new aspects, relevant to the comparison of synthetic profiles with observations carried out here.

As noted in the previous section, the approach presented in this paper is based on the analysis of the shapes of BELs and it does not explicitly incorporate the radial velocity curve modeling for candidates with a sequence of observed profiles. This is because our synthetic database presently contains profiles for only five equidistant orbital phases per SBHB configuration, whereas the spectroscopic searches with cadence of months to years correspond to a higher frequency of sampling of the SBHB radial velocity curves on average. 

The sparse sampling adopted in our synthetic database is a practical compromise motivated by considerations about its size.  Because of the extent of the parameter space, the number of sampled configurations quickly adds up to  about 42.5 million, even with a handful of choices per parameter. Note however that for promising SBHB candidates a denser coverage can be obtained for the sub-regions of the parameter space occupied by the binary configurations of interest. This includes a higher rate of sampling in orbital phase, so as to attempt to match the orbital phase of the observed SBHB candidates. This would provide a more stringent consistency check for the SBHB model by requiring that all observed profiles associated with a given SBHB candidate map into consistent values of $a$, $q$, $e$, $i$ (a requirement already used in this work), {\it and} that the time evolution of the profile shapes is consistent with the expected evolution of the orbital phase.

Another point worth noting is related to the inferred orbital separation of the candidate SBHBs. According to Figure~\ref{fig:BBH}, our method suggests that some of the candidates may be described by semimajor axes as small as $5000\,M$. Depending on the exact mass ratio and orbital eccentricity, these separations correspond to SBHs with mini-disks gravitationally truncated to a size of $\sim10^3\,M$ (see Figure~2 in paper~I), or about $\sim 10^{-3}-10^{-2}\,$pc for a binary with a mass of $10^{7-8}\,M_\odot$, respectively. By the time the mini-disks reach such compact sizes, their optical BLRs may be substantially truncated \citep[see work by][]{montuori11,montuori12}, resulting in the dimming of the broad optical emission lines considered in this work. A consequence of the BEL dimming would be a non-detection of some fraction of such compact SBHBs. Conversely, the same effect would be reflected in a more precipitous decrease of the PDFs of SBHB candidates with semimajor axes below $a\sim 10^4\,M$. The physics of binary BLRs is however not sufficiently understood in order to place firm constraints on their sizes (or photoionization properties for that matter). For this reason, we make no assumptions about the sizes of optical BLRs in binary mini-disks and attempt to circumvent the complexity by adopting the simplest of assumptions: if optical BLRs still exist in the most compact of SBHBs considered in our model, then their emission properties are set by the properties of the SBHB and the size of its mini-disks.

This work lays out an approach that can be used to estimate the SBHB parameters and their uncertainties once a sample of genuine sub-parsec binaries is available. A point worth reiterating however is that other physical processes can potentially mimic the emission signatures of SBHBs included in our database. These include but are not limited to recoiling SBHs \citep{blecha16} and local and global instabilities in single SBH accretion disks that can give rise to transient bright spots and spiral arms \citep{sbergmann03, lewis10, schimoia17}. In that sense, the model described in this paper can be used to interpret observed emission-line profiles in the context of the SBHB model but cannot be used to prove that they originate with genuine SBHB systems. 

\section{Conclusions}\label{sec:conclusions}

We present a method for comparison of the modeled and observed optical BELs, based on the principal component analysis, and use it to infer the properties of 88 SBHB candidates from the E12 spectroscopic survey. The new aspect of this method is that in addition to the parameter estimates it also provides a quantitative measure of the parameter degeneracy, thus allowing to establish the uncertainty intrinsic to such measurements. Our main results are as follows:


\begin{itemize}
\item We find that as a population, the observed SBHB candidates favor average value of the semimajor axis and standard deviation corresponding to $\log(a/M) \approx 4.20$ and 0.42, respectively, in agreement with expectations based on orbital evolution of SBHBs in circumbinary disks and the selection effects of spectroscopic surveys (see \S~\ref{sec:88candidates} and Figures~\ref{fig:meanstd} and \ref{fig:BBH}). They also favor configurations with comparable mass ratios, $q>0.5$, although this parameter suffers from a larger degree of degeneracy than $a$ (we provide the analytic fits to the 1D distribution functions of these parameters in equations~\ref{eqn:proba} and \ref{eqn:probq}). If the SBHB candidates analyzed here are shown to be true binaries, this result would suggest that there is a physical process that allows initially unequal mass systems to evolve toward comparable mass ratios (e.g., accretion that occurs preferentially onto the smaller of the SBHs). Alternatively, it would point to some, yet unspecified, selection bias that favors detection of higher $q$ binaries. Our method does not provide useful constraints on the orbital eccentricity, because this parameter suffers from a large degree of degeneracy.

\item SBHB candidates as a population show no strong preference for particular values of the angles that describe the orientation of the primary and secondary SBH mini-disks relative to the orbital plane. The most interesting implication of this is that binary candidates do not favor configurations in which the mini-disks are coplanar with the binary orbital plane more than any other configurations (see \S~\ref{sec:88candidates} and Figures~\ref{fig:individualt1t2} and \ref{fig:BBH}). If this is confirmed for true SBHBs, it would point to the presence of a physical mechanism which maintains the misalignment of the mini-disks (or causes them to be warped) down to sub-parsec binary separations. If so, the alignment of the mini-disks (and SBH spins if they are related to the mini-disk orientation) should evolve with binary separation. In this case, the orientation of the mini-disks inferred from observations would be an important indicator of whether the mechanism leading to the coplanar alignment is efficient, as predicted by some theoretical models.

\item  The inferred values of the optical depth parameter of the accretion disk wind in the two mini-disks cover a relatively wide range of values, $-3.4 \lesssim \log \tau_0 \lesssim 0.4$, and are moderately degenerate. Similarly, a majority of SBHB candidates in the E12 sample favor the values of $F_2/F_1 \lesssim 1$, and are consistent with the scenario in which the emission from the primary mini-disk either makes a dominant or a comparable contribution to the flux of the H$\beta$ emission line (see \S~\ref{sec:88candidates} and Figures~\ref{fig:meanstd} and \ref{fig:BBH}). An important implication of this result for spectroscopic searches for SBHBs is that they should consider the case in which a measured radial velocity curve is associated with the primary SBH, in addition to the commonly made assumption that they are associated with the secondary. The main difference between the two is that, all other things being the same, the former interpretation corresponds to more compact (and thus more strongly constrained) systems of SBHBs (see discussion in \S~\ref{sec:implications}). 

\item We find that epoch-to-epoch variability of the observed BELs provides an effective way to reduce the SBHB parameter degeneracies, as it helps to eliminate parameters that are not represented in all epochs of observations. Some of the strongest parameter constraints obtained with our method are achieved for individual SBHB candidates with many available observations (e.g., nine), thus providing an example of potential gains provided by continued spectroscopic monitoring (see \S~\ref{sec:individual}). 

\item In addition to the observed SBHB candidates, we perform the analysis of the spectra of a control group of AGNs and compare the two. The control AGNs favor similar average value of the semimajor axis, within the uncertainties, and exhibit a larger degree of degeneracy in this parameter (see \S~\ref{sec:individual} and Figures~\ref{fig:AGNmeanstd} and \ref{fig:Control}). The probability distributions for the remainder of the SBHB parameters look nearly the same for the two groups of objects. This similarity confirms that the approach presented here can be used to infer the parameters once a group of confirmed SBHBs is available, but cannot be used as a conclusive test of binarity. 
\end{itemize}

Further improvements to the presented method are possible by explicitly incorporating the modeling of the time-domain evolution of profile shapes into the model, at the expense of creating a larger database of synthetic profiles, with a higher sampling of the relevant SBHB parameters as a function of time. This would provide a more stringent consistency check by requiring that all observed profiles associated with a given SBHB candidate map into the consistent values of the binary parameters (a requirement already used in this work) {\it and} that the time-domain evolution of the profile shapes is consistent with the SBHB model. We defer this type of analysis to future work.

\acknowledgements
T.B. thanks the Kavli Institute for Theoretical Physics, where one portion of this work was completed, for its hospitality. 
This research was supported in part by the National Science Foundation under Grant No. NSF AST-1211677 (in the early stages of the project), Grant No. NSF PHY-1748958, and by the National Aeronautics and Space Administration under Grant No. NNX15AK84G issued through the Astrophysics Theory Program. Numerical simulations presented in this paper were performed using the high-performance computing cluster PACE, administered by the Office of Information and Technology at the Georgia Institute of Technology.


\appendix
\label{appendix}

\section{Computation of Principal Components}\label{sec:Calc}

In this section we briefly outline the approach used to calculate principal components and eigenprofiles for the profiles in our synthetic database. Each profile in the database is defined within the frequency range $(0.93, 1.07)\,\nu_0$ and is discretized using 600 equal frequency bins, where $\nu_0$ represents the rest frame frequency of the H$\beta$ emission line. We carry out the analysis in frequency space and only convert to wavelength space for visualization purposes (i.e., in figures). Moreover, all profiles are normalized in such a way that their maximum flux values are unity. 

\begin{figure}[t]
\centering
\includegraphics[width=0.6\textwidth, clip=true]{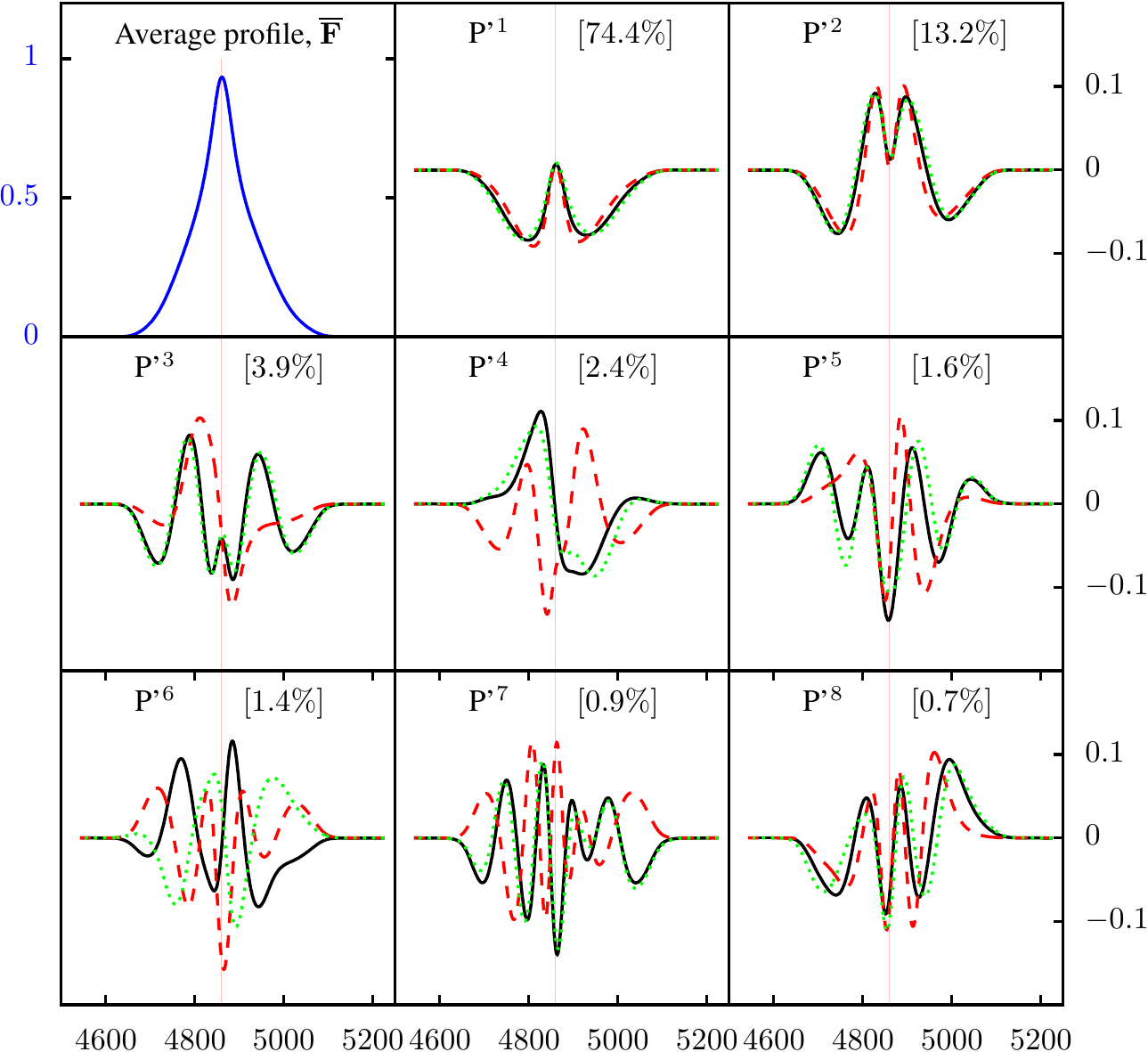}
\caption{The average profile (top left panel) and the first eight eigenprofiles (solid black lines) calculated for our synthetic database. The number in brackets marks the percentage of the database variance that corresponds to each eigenprofile. The red dashed (green dotted) line marks the eigenprofiles calculated for the portion of the synthetic profiles from circular (eccentric) SBHBs. The vertical red line marks the rest wavelength of the H$\beta$ line.}
\label{fig:eigenspectra}
\end{figure}

The database of all synthetic profiles can then be described as a matrix $\mathbf{F}$ of size $\left[\rm N \times M\right]$, where $\rm N \approx 4.25\times10^7$ rows represent the number of profiles and $\rm M=600$ columns represent the number of frequency bins. The average profile of the database can be calculated as a vector $\overline{\bf F}$ of size $\rm \left[1 \times M\right]$ with elements 
\begin{equation}
\overline{F}_m=\frac{1}{\rm N}\sum\limits_{n=1}^\mathrm{N} F^n_m \,\,.
\label{eq_average}
\end{equation}
The average profile of the synthetic database, shown in the top left panel of Figure~\ref{fig:eigenspectra}, is single peaked and fairly symmetric. All profiles in the synthetic database can be derived as a linear combination of the average profile and a finite number of eigenprofiles calculated for the synthetic database
\begin{equation}
F^n_m \approx \overline{F}_m + \sum\limits_{i=1}^{\textrm{I}} T^n_{i} P'^i_m \,\,,
\label{fig_decomp}
\end{equation}
where ``$\approx$" indicates that the profile reconstruction calculated in this way is an approximation of the actual profile due to truncation of the linear series in equation~\ref{fig_decomp}. Here, $\mathbf{T}$ is a matrix of size $\rm \left[N \times I\right]$ containing the principal components corresponding to each eigenprofile. In this work we choose $\rm I=20$ principal components, which is sufficient to precisely reconstruct all synthetic profiles in our database, and note that the choice $\rm I=M=600$ (the maximum possible value) does not improve the accuracy of reconstruction. The matrix $\mathbf{P}$ of size $\rm \left[M \times I\right]$ describes the set of eigenprofiles used to decompose the profiles in the modeled database and each column of this matrix represents one eigenprofile. Furthermore, the matrix $\mathbf{P}'$  with the size $ \rm \left[I \times M\right]$ is the transpose of $\mathbf{P}$.

The first-order eigenprofile, $\mathbf{P}_1$, is a unit vector pointing in the direction with the largest projected variance of the profile database.  Similarly, the $k^{\rm th}$-order eigenprofile, $\mathbf{P}_{k}$, is a unit vector pointing in the direction with the largest projected variance and is perpendicular to the $k-1$ lower-order eigenprofiles. For example,

\begin{align}
\mathbf{P}_1 =&\argmax_\mathbf{V:V'V=1} \left(\mathbf{V'X'XV}\right) \\
\vspace{0.75 cm}
\mathbf{P}_2 =&\argmax_\mathbf{V:V'V=1,V'P_1=0} \left(\mathbf{V'X'XV}\right)\,\,...
\end{align}
Here, $\mathbf{V}$ is a unit vector pointing in some arbitrary direction and $\mathbf{X}$ is a matrix with elements $X^n_m=F^n_m-\overline{F}_m$. $\mathbf{X'X}$ therefore represents the variance of the database and is proportional to its covariance matrix. $\mathbf{V'X'XV}$ represents the projection of the variance along the direction $\mathbf{V}$. The first-order principal axis, $\mathbf{P_1}$, is selected to be the unit vector $\mathbf{V}$ that maximizes the quantity $\mathbf{V'X'XV}$. The second-order principal axis, $\mathbf{P_2}$, points in a different direction $\mathbf{V}$ that maximizes $\mathbf{V'X'XV}$ and is perpendicular to $\mathbf{P_1}$. Note that here, the principal axes are the eigenvectors of the matrix $\mathbf{X'X}$, and the projected variances, $\mathbf{V'X'XV}$, are the eigenvalues.

Since $\mathbf{X'X}$ is positive semi-definite, its eigenvectors can be found by using the process of singular value decomposition. We use the following notation to describe this procedure
\begin{equation}
\mathbf{P}=\mathrm{svd}(\mathbf{X'X}) \,\,.
\end{equation}

The shapes of the first eight eigenprofiles, formally expressed as vectors $\mathbf{P'}^i$ with size $\rm \left[1 \times M\right]$, are shown in Figure~\ref{fig:eigenspectra} as black curves. For example, the top middle panel of Figure~\ref{fig:eigenspectra} shows that $\mathbf{P'}^1$ is an almost symmetric profile that accounts for 74.4\% of the variance of the database profiles, $\mathbf{P'}^2$ accounts for 13.2\%, and so on.

In our calculation, we divide the synthetic database into two portions, consisting of profiles calculated for the SBHBs on circular orbits (denoted by $^{\textrm{c}}\mathbf{X}$) and for the SBHBs on eccentric orbits ($^{\textrm{e}}\mathbf{X}$). We compute the covariance for each of the subsets, $^{\textrm{c}}\mathbf{X}$ and $^{\textrm{e}}\mathbf{X}$, before obtaining the total covariance by applying the partition relation
\begin{equation}
\label{eq:join}
\left(\mathbf{X'X}\right)^a_b=\left(^{\textrm{c}}\mathbf{X}'\,^{\textrm{c}}\mathbf{X}\right)^a_b + \left(^{\textrm{e}}\mathbf{X}'\,^{\textrm{e}}\mathbf{X}\right)^a_b + 
\frac{ ^{\textrm{c}} \textrm{N}\,^{\textrm{e}}\textrm{N}} { \left(^{\textrm{c}}\textrm{N} +\,^{\textrm{e}}\textrm{N}\right) } \left( ^{\textrm{c}} \overline{F}_a -\, ^{\textrm{e}} \overline{F}_a \right) \left( ^{\textrm{c}} \overline{F}_b -\, ^{\textrm{e}} \overline{F}_b \right) \,\,,
\end{equation}
where $a$ and $b$ are dummy indices of the variance matrix. Equation~\ref{eq:join} is numerically convenient because it allows parallel computing of $\mathbf{X'X}$. While we only divide the data into circular and eccentric SBHB cases (in order to compare them), one can in principle repeat this procedure for an arbitrary number of data subsets, making the analysis of large datasets more efficient. The parallelization can then be achieved by calculating the average profiles, $^{\textrm{c}} \overline{\bf F}$ and  $^{\textrm{e}} \overline{\bf F}$, as in equation~\ref{eq_average}, and by taking advantage of the property of the covariance matrix
\begin{equation}
\label{eq:covdef}
\left(\mathbf{X'X}\right)^a_b = \sum_{n=1}^{\textrm{N}} \left(F^n_a - \overline{F}_a \right) \left(F^n_b - \overline{F}_b \right) = 
\sum_{n=1}^{\textrm{N}} F^n_a \,F^n_b - \frac{1}{\textrm{N}}\sum_{n=1}^{\textrm{N}} F^n_a \, \sum_{n=1}^{\textrm{N}} F^n_b \,\,.
\end{equation}

Figure~\ref{fig:eigenspectra} shows that within the first 2 orders, the eigenprofiles for circular and eccentric cases are quite similar. The first significant difference between the two appears in the third order, where $^{\textrm{e}}\mathbf{P'}^3$ has a much more complex shape than $^{\textrm{c}}\mathbf{P'}^3$. This is in agreement with the finding reported in Paper~I that eccentric SBHBs can in principle produce more complex profiles due to a wider range of orbital velocities sampled by the orbiting binary. The eigenprofiles calculated for the entire database (including both circular and eccentric SBHBs) are more similar to those of the eccentric than circular SBHBs. This because our database contains a comparable number of eccentric cases ($\rm ^eN=24,546,000$) and circular cases ($\rm ^cN=17,816,400$).

\section{Results of analysis for 88 SBHB candidates}\label{sec:ParamTable}

We list the parameters inferred for 88 SBHB candidates from the E12 sample in Table~\ref{table:results} below. See the text for definitions and discussion of these parameters.

\startlongtable

\begin{deluxetable}{rccccccccccc}
\tablecaption{Inferred values of the binary parameters for the SBHB candidates from the E12 sample}\label{table:candidates}
\tablehead{\colhead{} & \colhead{SDSS} & \colhead{$\log(a/M)$} & \colhead{$\mathcal{S}_a$} & \colhead{$q$} & \colhead{$\mathcal{S}_q$} & \colhead{$\log \tau_0$} & \colhead{$\mathcal{S}_{\tau_0}$} & \colhead{ $\log(F_2/F_1)$ } &  \colhead{$\mathcal{S}_{\rm F_{2/1}} $} & \colhead{$\rm QI$} & \colhead{E12} 
}
\startdata
\input{Table2Data.txt}  
\enddata
\tablecomments{\,\,SDSS -- Name of the SBHB candidate. $\log(a/M)$ -- average value of $\log(a/M)$ with one standard deviation. $\mathcal{S}_a$ -- entropy associated with the value of $\log(a/M)$. $q$ -- average value of the mass ratio with one standard deviation. $\mathcal{S}_q$ -- entropy associated with the value of $q$. $\log \tau_0$ -- average value of $\log \tau_0$ with one standard deviation. $\mathcal{S}_{\tau_0}$ -- entropy associated with the value of $\log \tau_0$.
$\log(F_2/F_1)$ -- average value of $\log(F_2/F_1)$ with one standard deviation. $\mathcal{S}_{\rm F_{2/1}}$ -- entropy associated with the value of $\log(F_2/F_1)$. QI -- Quality index. E12 --  0 (1) corresponds to the SBHB candidates for which a statistically significant radial velocity modulation has not (has) been measured from observations.}
\label{table:results}
\end{deluxetable}

\onecolumngrid
\section{The effect of spectral noise on the values of the SBHB parameters}\label{sec:NoisyProf}

\begin{figure*}[t]
\centering
\includegraphics[width=0.48\textwidth, clip=true]{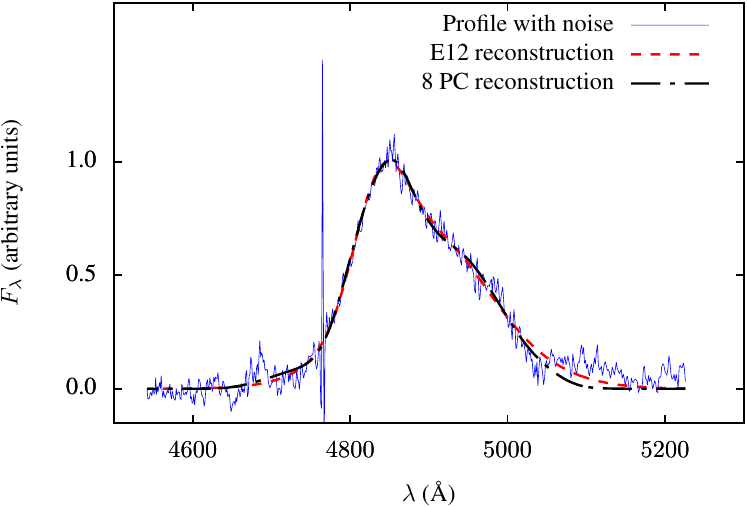}
\includegraphics[width=0.412\textwidth, trim = 30 0  0  0, clip=true]{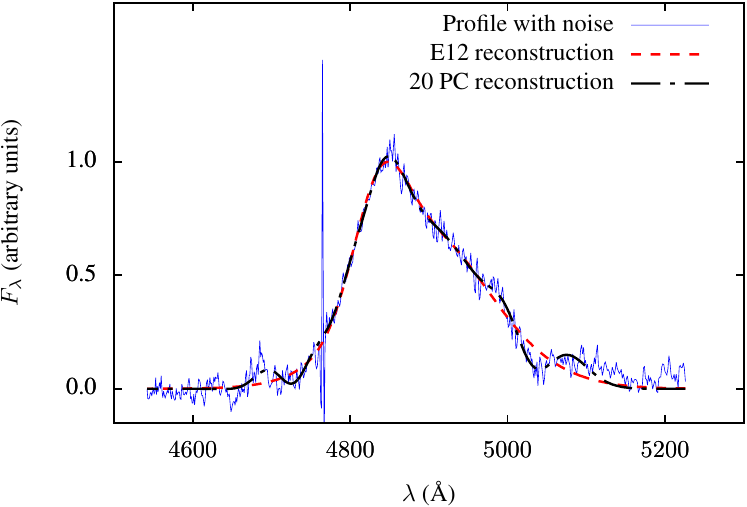} 
\caption{BEL profile obtained during the first epoch of observation of the SBHB candidate J093844 (blue solid line). The red dashed line is the parametric reconstruction of the BEL profile with noise by \citet{runnoe15}. The black dash-dot line in the left (right) panel is our reconstruction of the BEL profile with noise using 8 (20) principal components. A large spike noticeable leftward of 4800\,\AA\, is a spectral noise feature.}
\label{fig:NoisyRec}
\end{figure*}

As discussed in \S~\ref{sec:description}, the analysis described in this paper uses the parametric representation of the broad component of the H$\beta$ line from \citet{runnoe15}, which is obtained through spectral decomposition of the spectrum of SBHB candidates and AGNs from the E12 sample. The result of this procedure are the broad H$\beta$ profiles that are smooth (i.e., without visible contribution from the spectral noise), as illustrated in Figures~\ref{fig:cutoff} and \ref{fig:worstQI}. Working with the parametric reconstruction of the broad component of the H$\beta$ profiles is advantageous, because it allows us to make apple-to-apple comparisons with the synthetic  profiles produced by our model. This is because the synthetic broad H$\beta$ profiles also do not include any other emission components (i.e., quasar continuum, narrow H$\beta$ or [O~$\textsc{iii}$] lines) or noise by design. 

Because the modeling of the spectral noise is sidestepped in this approach, it does not allow us to assess the impact of the spectral noise on the inferred values of SBHB parameters. We examine this question here, by performing a case study of the H$\beta$ profiles that include the noise for a subset of SBHB candidates. Figure~\ref{fig:NoisyRec} shows an example of the H$\beta$ emission-line profile for one of the objects in the E12 sample. Specifically, it shows (a) the broad component of the H$\beta$ profile, deblended from the other emission components but including the spectral noise, (b) the smooth decomposition of this profile obtained by \citet{runnoe15} using two Gaussians, and (c) decomposition of the noisy profile obtained with either 8 or 20 principal components.

The reconstruction using Gaussians and 8 principal components trace each other closely and provide very similar representation of the broad H$\beta$ profile. The reconstruction using a full set of 20 principal components however departs from this representation and it also reproduces some features of the spectral noise. Therefore, have we used noisy broad H$\beta$ profiles from the very beginning, the uncertainty due to the spectral noise introduced in our analysis would be determined by such features. In order to quantify the resulting degree of uncertainty, we show the analysis of the noisy H$\beta$ profiles for the three SBHB candidates featured in Figure~\ref{fig:cutoff}: SDSS~093844, 095036 and 161911. 

\begin{figure*}[t]
\centering
\includegraphics[width=1.0\textwidth, clip=true]{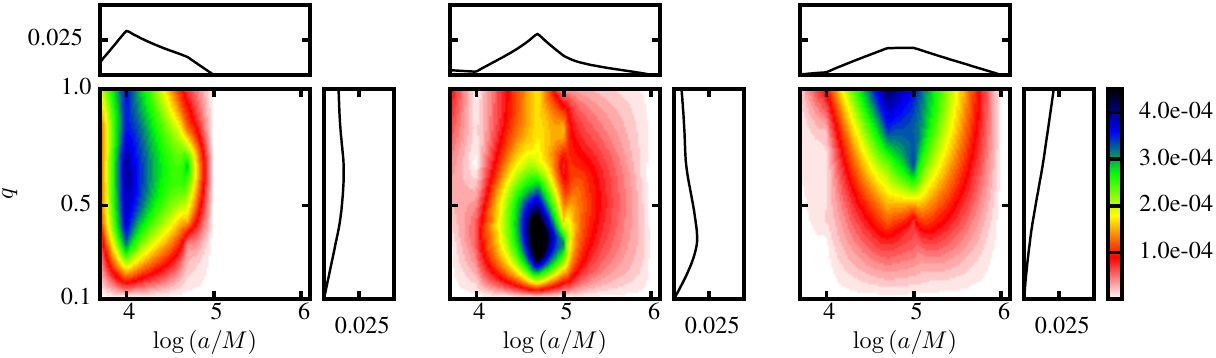}
\caption{2D probability density distribution in terms of $\log(a/M)$ and $q$ for SDSS~J093844 (left), J095036 (middle), and J161911 (right panel), inferred from the noisy H$\beta$ profiles. The rectangular insets show the 1D projections.}
\label{fig:individualaqNoise}
\end{figure*}

\begin{figure*}[t]
\centering
\includegraphics[width=1.0\textwidth, clip=true]{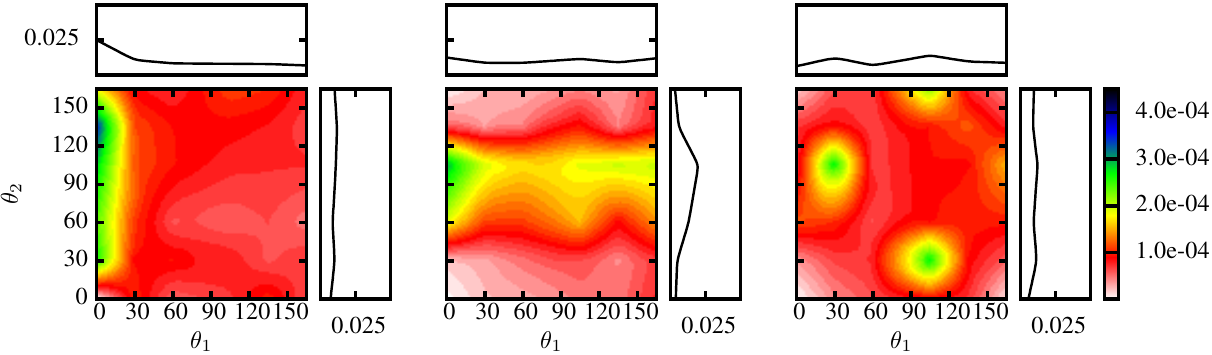}
\caption{2D probability density distribution in terms of $\theta_1$ and $\theta_2$ for SDSS~J093844 (left), J095036 (middle), and J161911 (right panel), inferred from the noisy H$\beta$ profiles. The rectangular insets show the 1D projections.}
\label{fig:individualt1t2Noise}
\end{figure*}

\begin{figure*}[t]
\centering
\includegraphics[width=1.0\textwidth, clip=true]{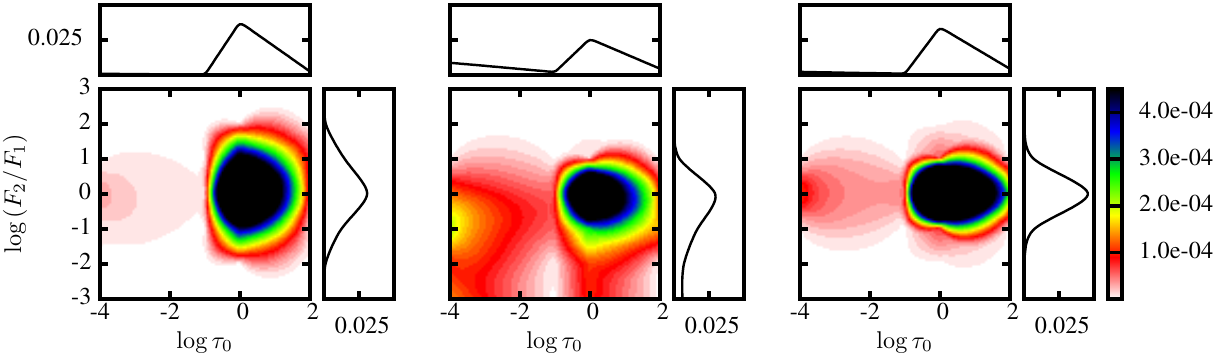}
\caption{2D probability density distribution in terms of $\log \tau_0$ and $\log{(F_2/F_1)}$ for SDSS~J093844 (left), J095036 (middle), and J161911 (right panel), inferred from the noisy H$\beta$ profiles. The rectangular insets show the 1D projections.}
\label{fig:individualtauF2F1Noise}
\end{figure*}

Figures~\ref{fig:individualaqNoise}, \ref{fig:individualt1t2Noise} and \ref{fig:individualtauF2F1Noise} show the 2D probability density distributions for the inferred SBHB parameters obtained in this way. These should be compared to Figures~\ref{fig:individualaq}, \ref{fig:individualt1t2} and \ref{fig:individualtauF2F1}, which are based on the analysis of smooth profiles obtained from the spectral decomposition by \citet{runnoe15}. Visual comparison of the two sets of PDFs shows that they are similar but not identical. Therefore, the presence of noise modifies the resulting probability distributions for SBHB parameters of individual objects. This effect is most visible in the 2D PDFs for $\log(a/M)$ and $q$ shown in Figure~\ref{fig:individualaqNoise} but is also noticeable in the remaining distributions.

%
%

Table~\ref{table:results2} summarizes the inferred values of the binary parameters for the three SBHB candidates, based on the analysis of the noisy broad H$\beta$ profiles using 20 principal components. These should be compared to the values shown for these candidates in Appendix~\ref{sec:ParamTable}. In all three cases the quality of the achieved match between the observed and synthetic profiles (indicated by the parameter QI) decreases. This is understandable, since reconstruction of the noise features introduces new bumps and indentations in the line profiles, which are not reproduced by our model. Comparison of the inferred values for $\log(a/M)$ and $q$ shows that they have not significantly changed, and that they remain well within the bounds of one standard deviation. The value of the parameter $\log \tau_0$ shows larger variation between the two analyses, albeit still within one standard deviation, which for this parameter is substantial. Similar is true for $\log(F_2/F_1)$. The impact of the spectral noise in the E12 dataset on the values of the inferred SBHB parameters is therefore smaller than the impact of the parameter degeneracy, represented by their standard deviation. 

\begin{figure}[]
\centering
\includegraphics[width=0.45\textwidth, clip=true]{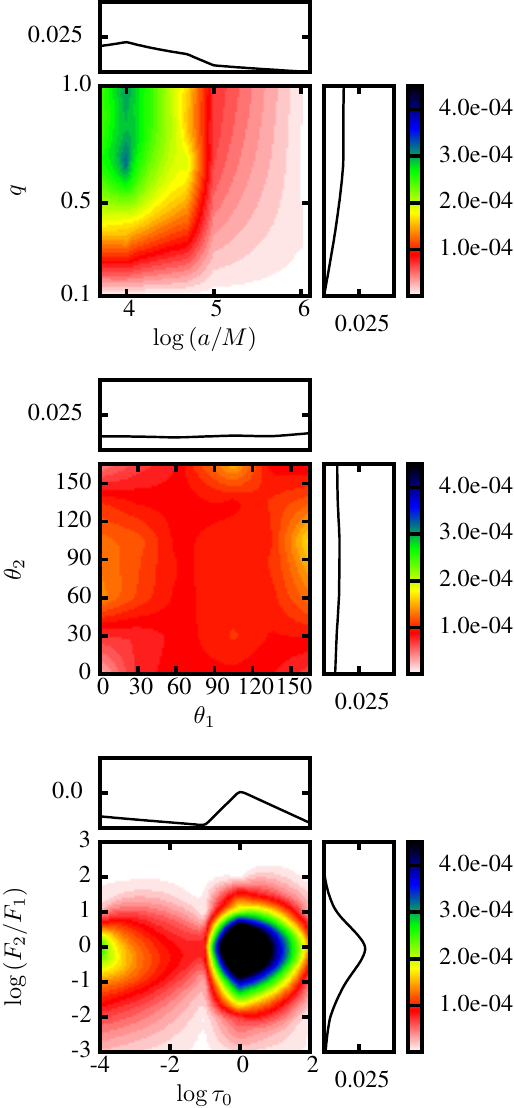}
\caption{2D probability density distributions in terms of $\log(a/M)$ and $q$ (top), $\theta_1$ and $\theta_2$ (middle), and $\log \tau_0$ and $\log{(F_2/F_1)}$ (bottom) for the 88 SBHB candidates from the E12 sample, inferred from the noisy H$\beta$ profiles. The rectangular insets show the 1D projections.}
\label{fig:BBH2}
\end{figure}

Finally, in Figure~\ref{fig:BBH2}, we show the resulting PDFs for the entire sample of 88 SBHB candidates from the E12 search, based on the analysis of the noisy broad H$\beta$ profiles. The distributions are calculated as the simple averages of distributions for individual candidates, and should be compared to those in Figure~\ref{fig:BBH}. The two sets of distributions appear very much alike, indicating that the spectral noise has a weak impact on the ability of this method to determine the average properties of the population of the SBHB candidates.

\begin{deluxetable*}{rccccccccccc}[t]
\tablecaption{Inferred values of the binary parameters for SDSS~093844, 095036 and 161911
}\label{table:candidates2}
\tablehead{\colhead{} & \colhead{SDSS} & \colhead{$\log(a/M)$} & \colhead{$\mathcal{S}_a$} & \colhead{$q$} & \colhead{$\mathcal{S}_q$} & \colhead{$\log \tau_0$} & \colhead{$\mathcal{S}_{\tau_0}$} & \colhead{ $\log(F_2/F_1)$ } &  \colhead{$\mathcal{S}_{\rm F_{2/1}} $} & \colhead{$\rm QI$} & \colhead{E12} 
}
\startdata
\input{Table3Data.txt}

\enddata
\tablecomments{\,\,SDSS -- Name of the SBHB candidate. $\log(a/M)$ -- average value of $\log(a/M)$ with one standard deviation. $\mathcal{S}_a$ -- entropy associated with the value of $\log(a/M)$. $q$ -- average value of the mass ratio with one standard deviation. $\mathcal{S}_q$ -- entropy associated with the value of $q$. $\log \tau_0$ -- average value of $\log \tau_0$ with one standard deviation. $\mathcal{S}_{\tau_0}$ -- entropy associated with the value of $\log \tau_0$.
$\log(F_2/F_1)$ -- average value of $\log(F_2/F_1)$ with one standard deviation. $\mathcal{S}_{\rm F_{2/1}}$ -- entropy associated with the value of $\log(F_2/F_1)$. QI -- Quality index. E12 --  0 (1) corresponds to the SBHB candidates for which a statistically significant radial velocity modulation has not (has) been measured from observations.}
\label{table:results2}
\end{deluxetable*}

\onecolumngrid

\clearpage
\bibliographystyle{aasjournal}
\bibliography{apj-jour,smbh}

\end{document}

%% file: Table2Data.txt
 1 & 001224 &       4.44 $\pm$       0.35 &       0.38 &       0.39 $\pm$       0.19 &       0.80 &      -2.81 $\pm$       1.64 &       0.43 &      -1.24 $\pm$       0.63 &       0.58 &      -0.09 & 1 \\ 
 2 & 002444 &       4.89 $\pm$       0.63 &       0.86 &       0.58 $\pm$       0.22 &       0.84 &      -0.26 $\pm$       1.27 &       0.78 &      -0.07 $\pm$       0.34 &       0.29 &       0.46 & 0 \\ 
 3 & 015530 &       4.05 $\pm$       0.30 &       0.55 &       0.75 $\pm$       0.22 &       0.83 &       0.27 $\pm$       0.81 &       0.41 &      -0.02 $\pm$       0.46 &       0.42 &       0.44 & 1 \\ 
 4 & 020011 &       4.17 $\pm$       0.35 &       0.62 &       0.66 $\pm$       0.24 &       0.89 &       0.21 $\pm$       0.75 &       0.36 &      -0.01 $\pm$       0.31 &       0.14 &       0.38 & 0 \\ 
 5 & 021259 &       4.77 $\pm$       0.22 &       0.49 &       0.64 $\pm$       0.16 &       0.71 &      -3.68 $\pm$       0.94 &       0.28 &      -0.42 $\pm$       0.26 &       0.36 &      -0.08 & 0 \\ 
 6 & 022014 &       3.88 $\pm$       0.14 &       0.42 &       0.68 $\pm$       0.23 &       0.89 &       0.18 $\pm$       0.76 &       0.33 &       0.23 $\pm$       0.50 &       0.49 &       0.52 & 0 \\ 
 7 & 031715 &       4.07 $\pm$       0.35 &       0.61 &       0.73 $\pm$       0.21 &       0.85 &      -0.26 $\pm$       1.26 &       0.41 &      -0.12 $\pm$       0.60 &       0.57 &       0.07 & 1 \\ 
 8 & 074007 &       4.35 $\pm$       0.43 &       0.77 &       0.73 $\pm$       0.22 &       0.85 &       0.20 $\pm$       0.86 &       0.35 &      -0.03 $\pm$       0.50 &       0.27 &       0.34 & 1 \\ 
 9 & 074157 &       4.34 $\pm$       0.44 &       0.79 &       0.71 $\pm$       0.22 &       0.83 &       0.05 $\pm$       1.11 &       0.53 &      -0.19 $\pm$       0.40 &       0.15 &       0.21 & 0 \\ 
10 & 075403 &       3.84 $\pm$       0.14 &       0.42 &       0.75 $\pm$       0.19 &       0.76 &      -0.28 $\pm$       1.09 &       0.38 &      -0.03 $\pm$       0.50 &       0.56 &       0.35 & 0 \\ 
11 & 080327 &       4.02 $\pm$       0.22 &       0.40 &       0.77 $\pm$       0.18 &       0.86 &      -0.88 $\pm$       1.52 &       0.75 &      -0.06 $\pm$       0.38 &       0.32 &       0.45 & 0 \\ 
12 & 081329 &       3.91 $\pm$       0.14 &       0.39 &       0.74 $\pm$       0.21 &       0.84 &      -0.28 $\pm$       1.46 &       0.67 &      -0.60 $\pm$       0.64 &       0.48 &       0.46 & 0 \\ 
13 & 082150 &       3.91 $\pm$       0.24 &       0.50 &       0.67 $\pm$       0.23 &       0.88 &       0.22 $\pm$       0.78 &       0.42 &       0.03 $\pm$       0.47 &       0.56 &       0.05 & 1 \\ 
14 & 082930 &       3.91 $\pm$       0.26 &       0.53 &       0.70 $\pm$       0.23 &       0.89 &       0.05 $\pm$       1.21 &       0.51 &      -0.18 $\pm$       0.51 &       0.53 &       0.22 & 0 \\ 
15 & 083223 &       4.29 $\pm$       0.44 &       0.77 &       0.79 $\pm$       0.19 &       0.78 &      -0.20 $\pm$       1.49 &       0.66 &      -0.09 $\pm$       0.24 &       0.09 &       0.66 & 0 \\ 
16 & 084313 &       4.88 $\pm$       0.53 &       0.77 &       0.70 $\pm$       0.22 &       0.89 &      -0.86 $\pm$       1.54 &       0.72 &      -0.06 $\pm$       0.34 &       0.34 &       0.46 & 0 \\ 
17 & 085431 &       4.74 $\pm$       0.27 &       0.55 &       0.69 $\pm$       0.23 &       0.88 &      -0.69 $\pm$       1.56 &       0.49 &      -0.08 $\pm$       0.32 &       0.25 &       0.59 & 0 \\ 
18 & 091833 &       3.73 $\pm$       0.09 &       0.20 &       0.37 $\pm$       0.15 &       0.64 &       0.16 $\pm$       0.59 &       0.14 &      -0.33 $\pm$       0.72 &       0.29 &      -0.55 & 0 \\ 
19 & 091928 &       3.84 $\pm$       0.14 &       0.41 &       0.81 $\pm$       0.20 &       0.76 &      -1.94 $\pm$       1.88 &       0.74 &      -0.18 $\pm$       0.47 &       0.40 &       0.26 & 1 \\ 
20 & 092712 &       3.93 $\pm$       0.14 &       0.39 &       0.41 $\pm$       0.18 &       0.77 &      -0.96 $\pm$       1.77 &       0.91 &      -0.80 $\pm$       0.68 &       0.70 &       0.17 & 1 \\ 
21 & 093100 &       3.85 $\pm$       0.12 &       0.32 &       0.68 $\pm$       0.19 &       0.80 &       0.21 $\pm$       1.10 &       0.62 &       0.14 $\pm$       0.26 &       0.14 &       0.43 & 0 \\ 
22 & 093653 &       4.17 $\pm$       0.35 &       0.68 &       0.67 $\pm$       0.21 &       0.87 &      -0.46 $\pm$       1.17 &       0.36 &      -0.13 $\pm$       0.40 &       0.39 &       0.11 & 0 \\ 
23 & 093844 &       4.19 $\pm$       0.34 &       0.67 &       0.65 $\pm$       0.22 &       0.88 &       0.21 $\pm$       0.65 &       0.27 &      -0.05 $\pm$       0.45 &       0.47 &       0.44 & 1 \\ 
24 & 094603 &       4.41 $\pm$       0.39 &       0.66 &       0.25 $\pm$       0.24 &       0.41 &      -2.73 $\pm$       1.62 &       0.47 &      -1.78 $\pm$       0.95 &       0.57 &      -0.71 & 1 \\ 
25 & 094620 &       4.68 $\pm$       0.21 &       0.40 &       0.57 $\pm$       0.20 &       0.89 &      -2.36 $\pm$       1.51 &       0.72 &      -0.49 $\pm$       0.36 &       0.34 &       0.35 & 1 \\ 
26 & 095036 &       4.64 $\pm$       0.28 &       0.44 &       0.43 $\pm$       0.17 &       0.44 &      -0.66 $\pm$       1.65 &       0.85 &      -0.47 $\pm$       0.51 &       0.46 &       0.59 & 1 \\ 
27 & 095539 &       4.17 $\pm$       0.38 &       0.59 &       0.74 $\pm$       0.22 &       0.83 &      -0.03 $\pm$       0.85 &       0.33 &      -0.17 $\pm$       0.58 &       0.55 &       0.16 & 1 \\ 
28 & 101438 &       4.01 $\pm$       0.30 &       0.56 &       0.65 $\pm$       0.24 &       0.91 &       0.31 $\pm$       0.80 &       0.40 &      -0.12 $\pm$       0.33 &       0.22 &       0.43 & 0 \\ 
29 & 102106 &       4.11 $\pm$       0.35 &       0.62 &       0.75 $\pm$       0.23 &       0.85 &      -0.15 $\pm$       1.42 &       0.61 &      -0.07 $\pm$       0.34 &       0.24 &       0.14 & 0 \\ 
30 & 102839 &       4.49 $\pm$       0.33 &       0.61 &       0.61 $\pm$       0.25 &       0.95 &      -0.52 $\pm$       1.76 &       0.82 &      -0.59 $\pm$       0.66 &       0.54 &       0.59 & 0 \\ 
31 & 104132 &       3.92 $\pm$       0.18 &       0.49 &       0.61 $\pm$       0.23 &       0.92 &       0.29 $\pm$       0.81 &       0.43 &      -0.05 $\pm$       0.36 &       0.29 &       0.56 & 0 \\ 
32 & 105041 &       3.97 $\pm$       0.09 &       0.16 &       0.11 $\pm$       0.06 &       0.17 &      -3.13 $\pm$       1.53 &       0.52 &      -2.22 $\pm$       0.36 &       0.40 &      -0.76 & 1 \\ 
33 & 105203 &       4.09 $\pm$       0.33 &       0.62 &       0.64 $\pm$       0.25 &       0.90 &      -0.69 $\pm$       1.66 &       0.83 &      -0.15 $\pm$       0.36 &       0.36 &       0.22 & 0 \\ 
34 & 110051 &       3.91 $\pm$       0.23 &       0.53 &       0.80 $\pm$       0.19 &       0.76 &      -0.70 $\pm$       1.72 &       0.75 &       0.00 $\pm$       0.23 &       0.09 &      -0.09 & 0 \\ 
35 & 110556 &       4.66 $\pm$       0.38 &       0.70 &       0.58 $\pm$       0.23 &       0.91 &       0.03 $\pm$       1.11 &       0.54 &      -0.26 $\pm$       0.34 &       0.28 &       0.72 & 1 \\ 
36 & 110742 &       3.85 $\pm$       0.14 &       0.42 &       0.69 $\pm$       0.21 &       0.82 &       0.29 $\pm$       0.73 &       0.36 &      -0.06 $\pm$       0.37 &       0.40 &       0.55 & 0 \\ 
37 & 111329 &       4.14 $\pm$       0.35 &       0.66 &       0.68 $\pm$       0.23 &       0.87 &      -0.25 $\pm$       1.16 &       0.37 &      -0.06 $\pm$       0.38 &       0.36 &       0.59 & 0 \\ 
38 & 111537 &       4.06 $\pm$       0.31 &       0.64 &       0.73 $\pm$       0.20 &       0.85 &      -0.49 $\pm$       1.64 &       0.58 &      -0.43 $\pm$       0.65 &       0.52 &      -0.35 & 1 \\ 
39 & 111916 &       4.23 $\pm$       0.43 &       0.79 &       0.69 $\pm$       0.24 &       0.89 &       0.16 $\pm$       0.91 &       0.39 &       0.19 $\pm$       0.51 &       0.52 &       0.22 & 0 \\ 
40 & 112751 &       4.49 $\pm$       0.40 &       0.66 &       0.75 $\pm$       0.20 &       0.76 &      -0.81 $\pm$       1.59 &       0.64 &      -0.02 $\pm$       0.37 &       0.36 &       0.40 & 0 \\ 
41 & 113330 &       4.25 $\pm$       0.39 &       0.63 &       0.61 $\pm$       0.23 &       0.85 &      -0.50 $\pm$       1.24 &       0.57 &      -0.31 $\pm$       0.33 &       0.39 &       0.44 & 1 \\ 
42 & 113651 &       4.07 $\pm$       0.34 &       0.63 &       0.71 $\pm$       0.23 &       0.87 &       0.17 $\pm$       0.78 &       0.30 &       0.05 $\pm$       0.46 &       0.33 &       0.31 & 0 \\ 
43 & 113706 &       3.89 $\pm$       0.18 &       0.50 &       0.70 $\pm$       0.21 &       0.83 &      -0.02 $\pm$       1.48 &       0.81 &      -0.44 $\pm$       0.56 &       0.52 &       0.57 & 0 \\ 
44 & 113904 &       3.79 $\pm$       0.14 &       0.36 &       0.55 $\pm$       0.22 &       0.84 &       0.41 $\pm$       0.90 &       0.40 &       0.19 $\pm$       0.39 &       0.43 &       0.02 & 1 \\ 
45 & 115158 &       4.39 $\pm$       0.48 &       0.64 &       0.65 $\pm$       0.23 &       0.90 &       0.02 $\pm$       0.96 &       0.60 &      -0.21 $\pm$       0.52 &       0.58 &       0.06 & 1 \\ 
46 & 115449 &       3.96 $\pm$       0.11 &       0.25 &       0.37 $\pm$       0.06 &       0.44 &      -3.23 $\pm$       1.39 &       0.54 &      -1.27 $\pm$       0.35 &       0.30 &      -0.11 & 0 \\ 
47 & 115644 &       4.36 $\pm$       0.43 &       0.73 &       0.71 $\pm$       0.24 &       0.88 &       0.17 $\pm$       0.85 &       0.41 &       0.10 $\pm$       0.43 &       0.37 &       0.44 & 0 \\ 
48 & 120924 &       3.86 $\pm$       0.17 &       0.49 &       0.59 $\pm$       0.22 &       0.88 &       0.24 $\pm$       0.72 &       0.44 &       0.31 $\pm$       0.37 &       0.44 &       0.36 & 1 \\ 
49 & 121113 &       4.22 $\pm$       0.32 &       0.64 &       0.66 $\pm$       0.23 &       0.89 &      -0.55 $\pm$       1.67 &       0.83 &      -0.32 $\pm$       0.42 &       0.36 &       0.48 & 0 \\ 
50 & 122811 &       4.00 $\pm$       0.26 &       0.56 &       0.72 $\pm$       0.22 &       0.86 &       0.04 $\pm$       1.17 &       0.44 &      -0.02 $\pm$       0.47 &       0.45 &       0.20 & 0 \\ 
51 & 123001 &       3.84 $\pm$       0.13 &       0.42 &       0.80 $\pm$       0.17 &       0.71 &       0.04 $\pm$       0.66 &       0.30 &      -0.05 $\pm$       0.76 &       0.59 &       0.35 & 0 \\ 
52 & 124551 &       4.71 $\pm$       0.27 &       0.54 &       0.75 $\pm$       0.20 &       0.78 &       0.04 $\pm$       0.77 &       0.27 &      -0.22 $\pm$       0.38 &       0.26 &       0.55 & 0 \\ 
53 & 125142 &       3.96 $\pm$       0.21 &       0.48 &       0.65 $\pm$       0.23 &       0.88 &       0.19 $\pm$       1.31 &       0.77 &      -0.46 $\pm$       0.55 &       0.53 &       0.47 & 1 \\ 
54 & 125809 &       3.71 $\pm$       0.04 &       0.04 &       0.79 $\pm$       0.15 &       0.63 &      -1.22 $\pm$       1.82 &       0.47 &      -0.40 $\pm$       0.52 &       0.06 &      -0.20 & 0 \\ 
55 & 130534 &       4.59 $\pm$       0.42 &       0.76 &       0.81 $\pm$       0.18 &       0.73 &      -1.07 $\pm$       1.86 &       0.65 &      -0.26 $\pm$       0.53 &       0.49 &      -0.01 & 0 \\ 
56 & 131945 &       4.70 $\pm$       0.00 &       0.00 &       0.83 $\pm$       0.15 &       0.34 &       0.00 $\pm$       1.21 &       0.55 &       0.06 $\pm$       0.65 &       0.57 &       0.24 & 1 \\ 
57 & 132704 &       4.10 $\pm$       0.35 &       0.60 &       0.66 $\pm$       0.23 &       0.88 &      -0.22 $\pm$       1.25 &       0.48 &      -0.11 $\pm$       0.42 &       0.15 &      -0.27 & 0 \\ 
58 & 133432 &       3.91 $\pm$       0.13 &       0.39 &       0.77 $\pm$       0.19 &       0.76 &      -0.84 $\pm$       0.96 &       0.67 &      -0.29 $\pm$       0.56 &       0.56 &       0.51 & 0 \\ 
59 & 134617 &       3.89 $\pm$       0.14 &       0.42 &       0.68 $\pm$       0.21 &       0.89 &      -0.63 $\pm$       1.71 &       0.76 &      -0.78 $\pm$       0.65 &       0.51 &       0.49 & 0 \\ 
60 & 140007 &       4.72 $\pm$       0.32 &       0.64 &       0.74 $\pm$       0.19 &       0.83 &      -1.11 $\pm$       0.90 &       0.42 &      -0.07 $\pm$       0.36 &       0.33 &       0.37 & 0 \\ 
61 & 140251 &       4.11 $\pm$       0.37 &       0.68 &       0.66 $\pm$       0.24 &       0.91 &       0.15 $\pm$       0.65 &       0.27 &       0.21 $\pm$       0.39 &       0.34 &       0.26 & 1 \\ 
62 & 140700 &       3.82 $\pm$       0.14 &       0.42 &       0.76 $\pm$       0.20 &       0.83 &      -0.01 $\pm$       0.37 &       0.08 &      -0.03 $\pm$       0.64 &       0.61 &       0.33 & 0 \\ 
63 & 141213 &       3.77 $\pm$       0.10 &       0.41 &       0.74 $\pm$       0.19 &       0.78 &      -0.78 $\pm$       1.16 &       0.26 &      -0.00 $\pm$       0.63 &       0.44 &       0.16 & 0 \\ 
64 & 141300 &       4.09 $\pm$       0.35 &       0.66 &       0.72 $\pm$       0.23 &       0.87 &       0.18 $\pm$       1.01 &       0.42 &      -0.03 $\pm$       0.31 &       0.20 &       0.40 & 0 \\ 
65 & 143123 &       3.92 $\pm$       0.24 &       0.61 &       0.62 $\pm$       0.19 &       0.83 &       0.27 $\pm$       0.79 &       0.34 &       0.04 $\pm$       0.28 &       0.29 &       0.42 & 0 \\ 
66 & 143455 &       3.93 $\pm$       0.09 &       0.15 &       0.52 $\pm$       0.19 &       0.88 &       0.17 $\pm$       1.03 &       0.33 &       0.12 $\pm$       0.29 &       0.43 &       0.44 & 0 \\ 
67 & 151132 &       4.47 $\pm$       0.47 &       0.82 &       0.80 $\pm$       0.18 &       0.74 &      -0.81 $\pm$       1.68 &       0.66 &      -0.39 $\pm$       0.53 &       0.51 &       0.24 & 1 \\ 
68 & 151443 &       4.35 $\pm$       0.33 &       0.62 &       0.70 $\pm$       0.25 &       0.93 &       0.16 $\pm$       0.64 &       0.10 &      -0.17 $\pm$       0.48 &       0.40 &       0.65 & 0 \\ 
69 & 152316 &       3.91 $\pm$       0.16 &       0.40 &       0.80 $\pm$       0.19 &       0.77 &       0.41 $\pm$       0.87 &       0.43 &       0.06 $\pm$       0.67 &       0.53 &       0.46 & 0 \\ 
70 & 152939 &       4.14 $\pm$       0.42 &       0.65 &       0.72 $\pm$       0.22 &       0.82 &       0.11 $\pm$       0.92 &       0.47 &      -0.03 $\pm$       0.45 &       0.45 &       0.22 & 0 \\ 
71 & 152942 &       3.88 $\pm$       0.14 &       0.40 &       0.79 $\pm$       0.16 &       0.73 &      -0.09 $\pm$       0.95 &       0.31 &      -0.33 $\pm$       0.79 &       0.62 &       0.32 & 0 \\ 
72 & 153434 &       4.61 $\pm$       0.38 &       0.74 &       0.77 $\pm$       0.19 &       0.80 &      -0.10 $\pm$       1.21 &       0.51 &      -0.03 $\pm$       0.35 &       0.26 &       0.38 & 0 \\ 
73 & 153636 &       3.82 $\pm$       0.12 &       0.27 &       0.46 $\pm$       0.14 &       0.65 &      -3.40 $\pm$       1.21 &       0.76 &      -1.02 $\pm$       0.50 &       0.45 &      -1.08 & 1 \\ 
74 & 153644 &       4.28 $\pm$       0.45 &       0.85 &       0.77 $\pm$       0.21 &       0.81 &       0.04 $\pm$       1.09 &       0.49 &      -0.05 $\pm$       0.40 &       0.37 &      -0.12 & 1 \\ 
75 & 154340 &       4.32 $\pm$       0.43 &       0.76 &       0.79 $\pm$       0.20 &       0.79 &       0.00 $\pm$       1.12 &       0.45 &      -0.06 $\pm$       0.55 &       0.50 &       0.34 & 0 \\ 
76 & 154637 &       3.89 $\pm$       0.14 &       0.42 &       0.73 $\pm$       0.20 &       0.86 &      -1.40 $\pm$       1.67 &       0.73 &      -0.15 $\pm$       0.55 &       0.49 &       0.15 & 1 \\ 
77 & 155654 &       3.95 $\pm$       0.11 &       0.12 &       0.43 $\pm$       0.10 &       0.57 &      -1.51 $\pm$       1.36 &       0.67 &      -0.49 $\pm$       0.28 &       0.38 &       0.15 & 1 \\ 
78 & 160243 &       4.04 $\pm$       0.31 &       0.71 &       0.72 $\pm$       0.22 &       0.84 &       0.02 $\pm$       0.91 &       0.46 &       0.06 $\pm$       0.31 &       0.17 &       0.25 & 0 \\ 
79 & 160536 &       3.99 $\pm$       0.25 &       0.51 &       0.73 $\pm$       0.23 &       0.86 &       0.37 $\pm$       0.88 &       0.41 &       0.03 $\pm$       0.29 &       0.21 &       0.41 & 0 \\ 
80 & 161911 &       4.81 $\pm$       0.26 &       0.50 &       0.72 $\pm$       0.22 &       0.89 &       0.02 $\pm$       1.02 &       0.42 &      -0.08 $\pm$       0.36 &       0.32 &       0.69 & 1 \\ 
81 & 162914 &       3.91 $\pm$       0.31 &       0.60 &       0.78 $\pm$       0.21 &       0.80 &       0.04 $\pm$       1.04 &       0.43 &      -0.01 $\pm$       0.24 &       0.08 &       0.11 & 1 \\ 
82 & 163020 &       3.78 $\pm$       0.12 &       0.42 &       0.63 $\pm$       0.24 &       0.90 &       0.02 $\pm$       0.80 &       0.37 &       0.13 $\pm$       0.55 &       0.62 &      -0.19 & 0 \\ 
83 & 165118 &       4.41 $\pm$       0.28 &       0.65 &       0.53 $\pm$       0.20 &       0.82 &       0.09 $\pm$       0.44 &       0.14 &      -0.15 $\pm$       0.43 &       0.41 &       0.51 & 0 \\ 
84 & 165255 &       4.07 $\pm$       0.32 &       0.58 &       0.65 $\pm$       0.24 &       0.90 &       0.26 $\pm$       0.70 &       0.32 &      -0.08 $\pm$       0.36 &       0.32 &       0.66 & 0 \\ 
85 & 170341 &       4.17 $\pm$       0.24 &       0.44 &       0.42 $\pm$       0.22 &       0.73 &      -2.34 $\pm$       1.78 &       0.29 &      -1.10 $\pm$       0.61 &       0.53 &       0.07 & 0 \\ 
86 & 171448 &       3.82 $\pm$       0.15 &       0.42 &       0.71 $\pm$       0.23 &       0.88 &       0.06 $\pm$       1.33 &       0.64 &       0.12 $\pm$       0.63 &       0.54 &      -0.45 & 0 \\ 
87 & 172711 &       4.19 $\pm$       0.35 &       0.61 &       0.63 $\pm$       0.24 &       0.89 &       0.05 $\pm$       1.07 &       0.44 &      -0.27 $\pm$       0.45 &       0.43 &       0.68 & 0 \\ 
88 & 180545 &       4.23 $\pm$       0.35 &       0.64 &       0.69 $\pm$       0.23 &       0.89 &       0.20 $\pm$       0.98 &       0.42 &       0.04 $\pm$       0.36 &       0.26 &       0.25 & 0 \\ 

%% file: Table3Data.txt
23 & 093844 &       4.10 $\pm$       0.31 &       0.59 &       0.68 $\pm$       0.23 &       0.89 &       0.02 $\pm$       1.04 &       0.25 &      -0.06 $\pm$       0.48 &       0.50 &       0.06 & 1 \\ 
26 & 095036 &       4.65 $\pm$       0.39 &       0.62 &       0.52 $\pm$       0.22 &       0.83 &      -1.15 $\pm$       1.86 &       0.73 &      -0.94 $\pm$       0.87 &       0.59 &      -0.23 & 1 \\ 
80 & 161911 &       4.80 $\pm$       0.25 &       0.57 &       0.78 $\pm$       0.20 &       0.80 &      -0.12 $\pm$       1.23 &       0.45 &      -0.01 $\pm$       0.26 &       0.14 &       0.29 & 1 \\